\documentclass[12pt,a4paper,final]{iopart}

\usepackage{iopams}  
\usepackage{setstack}
\usepackage{iopams_add}
\usepackage{graphicx}
\usepackage[breaklinks=true,colorlinks=true,linkcolor=blue,urlcolor=blue,citecolor=blue]{hyperref}
\usepackage{latexsym}
\usepackage{graphics}
\usepackage{graphicx}
\usepackage{color}
\newcommand{\be}{\begin{equation}}
\newcommand{\ee}{\end{equation}}
\newcommand{\ba}{\begin{eqnarray}}
\newcommand{\ea}{\end{eqnarray}}
\newcommand{\ar}{\arrowvert}

\newcommand{\da}{\dagger}

\newcommand{\Imag}{\mathop{\mathrm{Im}}}
\newcommand{\Real}{\mathop{\mathrm{Re}}}

\begin{document}
\title{Light ``Higgs'', yet strong interactions}

\author{Rafael L. Delgado, Antonio Dobado \\ and Felipe J. Llanes-Estrada}
\address{Departamento de F\'isica Te\'orica I, Universidad Complutense de Madrid, 28040 Madrid, Spain}
\ead{dobado@fis.ucm.es}

\begin{abstract}
The claimed finding of a light Higgs boson makes the minimal Standard Model unitary. 
Yet we recall that the general low-energy dynamics for the minimal electroweak symmetry breaking sector with three Goldstone bosons and one light scalar is not so.  We construct the effective Lagrangian for these four particles and their scattering amplitudes, that can be extracted from LHC experiments when longitudinal $W$, $Z$ modes be properly isolated for $E\gg M_W$ (Equivalence Theorem).
We then  observe the known increase in interaction strength with energy and explore various unitarization methods in the literature in the absence of other new physics (as LHC experiments fail to report anything up to 600 GeV).
Our generic conclusion is that
for most of parameter space the high energy scattering of the longitudinal $W$'s is strongly interacting  (with the Minimal Standard Model a remarkable exception).   
We find and study a second $\sigma$-like scalar pole of the $W_L W_L$ amplitude.
\end{abstract}


\maketitle
\section{Introduction}

Both CMS~\cite{CMS} and ATLAS~\cite{ATLAS} have reported an excess of events in four-lepton (largely 4$\mu$) spectra compatible with a Minimal Standard Model (MSM) Higgs boson, with a claimed statistical significance of signal over background around 5 $\sigma$. The two collaborations have also reported possible excesses in the two-photon~\cite{twophotons} and $WW$~\cite{Aad:2013wqa,CMS:aya} channels, though these are less clear to us due to the uncalculated backgrounds. 
There are various other hints and claims by the same and other collaborations that are at too early a stage to draw definite conclusions. 

Most importantly, the examination of numerous channels has revealed no evidence of additional new physics up to 600 GeV, although searches continue~\cite{searches}.

If the situation settles in the existence of only one Higgs-like boson at low energies, then the Electroweak Symmetry Breaking Sector (EWSBS) of the Standard Model (SM) is composed of this  scalar resonance and the three would-be Goldstone bosons (WBGB). With help of the global symmetry breaking pattern
$SU(2)_L \times SU(2)_R \to SU(2)_C$ it is easy to write down the effective Electroweak Chiral Lagrangian (EWChL)~\cite{Appelquist} for the three WBGB's 
 in a similar way as it is done in standard Chiral Perturbation Theory (ChPT) for pions~\cite{ChPT}. 
At energies sufficiently higher than $100$ GeV (their typical mass scale) we can identify them with the longitudinal components of the $W$ and $Z$ bosons, thanks to the Equivalence Theorem~\cite{ET}, which allows to experimentally access them. The separation of the transverse and longitudinal modes of the $W$'s now seems very promising with LHC data~\cite{belyaev}.
 
We then couple the recently discovered new scalar  $\varphi$ to the WBGB  in a standard way and thus write down the most general effective Lagrangian describing the 
low energy dynamics of  these four modes~\cite{Grinstein:2007iv,scalar} in section~\ref{Lagrangian}. It is interesting to realize that the fact that we have discovered four scalar light modes with a strong gap (no further states until at least about 600 GeV) strongly suggests that we should take seriously the possibility that these scalars are to be interpreted as the Goldstone Bosons  (GB) corresponding to some spontaneous symmetry breaking happening in the EWSBS at some higher scale $f$~\cite{GB}.

In the main part of this article, we first concentrate on
the ``high'' energy behavior of the scattering amplitudes for the
longitudinal components of the electroweak gauge bosons (EWGB) or equivalently WBGB's. Here, we mean $M_W$, $M_\varphi=\Or(100{\rm GeV})\ll E$ but we will consider at most $E\simeq \Lambda  \simeq 4\pi v\simeq 3$ TeV since the validity of effective approaches is not granted at still higher energies.

The perturbative amplitudes are given in section~\ref{sec:scattering}, as are also the partial wave expansions and the coupled-channel formalism connecting the $W_LW_L$ and $\varphi\varphi$ channels. The low-energy Lagrangian contains two parameters, $v=246$ GeV (that breaks electroweak symmetry) and $f$ which typically represents a higher scale still undetermined. In the MSM $f=v$ and only one parameter is necessary to describe the interaction. In this case, the Higgs saturates unitarity; but if $f\ne v$, the elastic scattering WBGB amplitude, proportional to $s= E_{cm}^2$, eventually grows out of the unitarity bound. Then, the perturbative description looses its validity as the interactions become strong.
In addition, there is a prominent inelastic reaction $W_LW_L\to \varphi\varphi$ controlled by two additional parameters (pure numbers, denoted $\alpha$ and $\beta$ below) and the pure scalar potential with parameters $\lambda_3$, $\lambda_4$. If $\alpha^2=\beta $ the channels decouple, otherwise their coupling also grows with the square of the energy.

We then use different techniques to unitarize
the effective low-energy amplitudes in a physically sensible way to higher energies in section~\ref{sec:unitarity}. We do not commit to a specific microscopic model~\cite{Pich:2013fba}
of the high energy theory (such as Composite Higgs,
Walking Technicolor (see for example the second article in reference~\cite{scalar} and references therein),
an electroweak scale Dilaton~\cite{Grinstein} or others, but instead employ generally valid approaches such as dispersion relations~\cite{Truong,Dobado} and on-shell factorization~\cite{Oller} (algebraic unitarization formulae). 

Our conclusion is that, safe the exception of the MSM, the scattering amplitudes become characteristically strong at the TeV scale and unitarization is needed. When projecting over $s$-wave and analytically continuing to the second Riemann sheet we find a broad pole, very reminiscent of the renowned scalar $\sigma$ that provides for central attraction of the nuclear potential in QCD, with several numerical computations shown in section~\ref{sec:numeric}. We discuss whether one can or not think of it as an additional particle, and if so why is it interesting, 
in section~\ref{sec:conclusions}, where we also wrap our discussion.

\section{Generic effective Lagrangian} \label{Lagrangian}

In the spirit of the previous discussion, we adopt the most general Lagrangian describing the low energy dynamics of the four 
 light modes (three $w$ WBGB and the Higgs $\varphi$). 
This can be written as a $SU(2)_L \times SU(2)_R/SU(2)_C = SU(2) \simeq S^3$ Non-linear Sigma Model (NLSM) coupled to a scalar field $\varphi$ in leading order of the derivative, chiral expansion, as
\be \label{genericLagrangian}
{\cal L}=\frac{v^2}{4}g(\varphi/f)Tr(D_\mu U)^\dag
D^\mu U+\frac{1}{2}\partial_\mu \varphi \partial^\mu
\varphi-V(\varphi) \ee where $U$ is a field taking values in the
$SU(2)$ coset and that we will parametrize as $U=\sqrt{1-\tilde
\omega^2/v^2}+ i \tilde  \omega/v$, with $\tilde \omega =
\omega_a\tau^a$ the WBGB; $D_\mu U=\partial_\mu U + W_\mu U- U  Y_\mu$,
$W_\mu = - g i W_\mu^i \tau^i/2$, $Y_\mu = - g' i B_\mu^i \tau^3/2$ coupling
the transverse gauge bosons;
$v = 246$ GeV the MSM Higgs-doublet vacuum expectation value (or in terms of Fermi's weak constant, $v^2=1/(\sqrt{2}G_F)$; $f$ is an arbitrary, new, dynamical energy scale needed for  the generic dynamics of the EWSBS;  $g(x)$ is an arbitrary
analytical functional of the scalar field  
\be \label{gexpansion}
g(\varphi/f)= 1 + \sum _{n=1}^{\infty}g_n
\left(\frac{\varphi}{f}\right)^n= 1 +2 \alpha \frac{\varphi}{f} +\beta\left(\frac{\varphi}{f}\right)^2+..
\ee 
where only the first terms are relevant for this work and we have parametrized them in terms of two arbitrary $\alpha$ and $\beta$ real constants instead of the more common $a$ and $b$ in~\cite{scalar}. This is because we are using the scale $f$ instead of $v$ to normalize the scalar field $\varphi$, which we consider more appropriate for our work here. Obviously we have $a= \alpha v/f$ and
$b= \beta v^ 2/ f^ 2$. 

If we accept at face value the $WW$ data from CMS and ATLAS~\cite{Aad:2013wqa,CMS:aya} (while also keeping in mind various experimental caveats~\footnote{Very strong background subtractions are needed, and the background functions are not known from first principles. Instead the backgrounds are fitted in off mass-peak control regions and then extrapolated to the signal zone of 125 GeV employing Monte Carlo simulations. The ``transfer factors'' for these extrapolations depend on parton distribution functions, QCD corrections, and Monte Carlo model choices. ATLAS posits an error varying from only 2\% for no-jet $WW$ events to $O(50\%)$ for events with two or more jets. In addition, the zero-width approximation is adopted for the Higgs and the Standard Model tensor structure is assumed in the analysis.}
), the parameter $f/\alpha$ (equivalently, $a$) is constrained to 95\% confidence level (2$\sigma$) to lie within the respective experimental bands
\ba \label{boundexp1}
\frac{f}{\alpha} \in (225,350){\rm GeV}  \ \ {\rm or}\  \  a \in (0.70,1.1) \ \ {\rm (CMS)} \\ 
\label{boundexp2}
\frac{f}{\alpha} \in (185,285){\rm GeV} \ \ {\rm or}\  \ a \in (0.87,1.3) \ \ {\rm (ATLAS)}\ .
\ea
These can be extracted from figures 7b and 13 of \cite{CMS:aya} and \cite{Aad:2013wqa} respectively, that bind the ratio between the $hWW$ coupling quotiented by its value in the SM, or $\kappa_v$, so that $f/\alpha= v/\kappa_v$, assuming other parameters are kept fixed at their SM value.
Nevertheless, in our numeric computations below we will explore a broader range of values of $f/\alpha$ in case these bounds are relaxed by later, more accurate data, and because of the intrinsic theoretical interest of the calculation.
No relevant (order O(1))  constraint on $\beta$ is known to us, because no two-Higgs final state has yet been detected.

In addition some authors have combined the results of the two experiments (in spite of the different systematics) or made theory-assisted analysis that can be found in
 \cite{bounds}. The possible bounds on the $a$ and $b$ parameters, or alternatively $f$ and $b$, are then somewhat different.
It is also worth mentioning that an analysis of the LEP precision observables~\cite{Azatov:2012bz} (though dependent on a loop-cutoff parameter that entails a large degree of arbitrariness) is consistent with Eq.~(\ref{boundexp1}) and (\ref{boundexp2}), yielding (assuming a positive $a$-parameter)
\be
\frac{f}{\alpha} \in (205,270){\rm GeV}  \ \ {\rm or}\  \  a\in(0.91,1.2)\ \ {\rm LEP}
\ee
at 99\% confidence level.

Finally, $V$ is an arbitrary analytical potential for the scalar field,
\ba \label{potential}
V(\varphi) = \sum _{n=0}^{\infty}V_n \varphi^n
\equiv  V_0 + 
\frac{M_\varphi^2}{2} \varphi^2 + \sum _{n=3}^{\infty} 
\lambda_n \varphi^n 
\ea
in obvious notation.

The Lagrangian written in this simple, canonical form of Eq.~(\ref{genericLagrangian}),
encodes the low energy dynamics of any model fulfilling the simple 
hypothesis of having only four light modes coming from the EWSBS, in terms of a few parameters. 

The most obvious and paradigmatic example is simply the MSM~\cite{GWS} which can be obtained  by just setting $f=v$, $\alpha=\beta=1$, $g_i=0$ for $i\ge3$ and by identifying the Higgs field $H$  with
$\varphi$ so that $M_H^2=M_\varphi^2= 2\lambda v^2$, and the scalar self-couplings as $\lambda_3= \lambda v$, $\lambda_4 = \lambda/4$ and $\lambda_i=0$ for $i \ge 4$. The function $g$ is given by
\be
g_{\rm MSM}(\varphi/v)=\left(1+\frac{\varphi}{v}\right)^2 \ .
\ee 

A second, very interesting, family of models includes those in which the scalar resonance is understood  as the Goldstone boson associated to the
spontaneous breaking  of the approximate scale (and conformal)
invariance of the unknown symmetry breaking sector, i.e., those that identify the light scalar as a dilaton~\cite{Grinstein}. 
The $g$ function is similar to that of the MSM but with the new scale $f$ playing the role of $v$
\be
g_{\rm dilaton}(\varphi/f)=\left(1+\frac{\varphi}{f}\right)^2
\ee
and in its simplest version
the potential is given by: 
\be
V(\varphi)=\frac{M_\varphi^2}{4 f^2}(\varphi+f)^2\left[\log \left(1 + \frac{\varphi}{f}\right)-
\frac{1}{4}\right]
\ee but more complicated potentials are also possible, depending on the precise form of the breaking of conformal invariance. Obviously in these dilaton models one has $\alpha=\beta=1$ and in general $f \ne v$.

A third type of model that can be treated with
Eq.~(\ref{genericLagrangian}) is that in which the WBGB and the scalar are
considered as GB associated to some spontaneous global symmetry
breaking. For example one possibility is to consider the coset
$SO(5)/SO(4)=S^4$  ~\cite{SO(5)} coming from a symmetry breaking from $SO(5)$ to $SO(4)$ at some scale $f$, which provides exactly four GB, followed by a second breaking from $SO(4)$ to $SO(3)$ at the electroweak scale $v$. This is typical of Composite Higgs Models in 4D or with extra dimensions, and in this case 
\be
g_{\rm CHM}(\varphi/f)=\frac{f^2}{v^2}\sin^2\left(\theta +\frac{\varphi}{f}\right)
\ee 
where $\sin \theta = v/f$, and various composite Higgs models can be selected by appropriately choosing the potential $V(\varphi)$.

The Lagrangian density in Eq.~(\ref{genericLagrangian})
provides the scattering amplitudes among the four light particles.
The high-energy $(s\gg M_\varphi^2)$ scattering of the WBGB is related 
to the scattering of the longitudinal components of the electroweak bosons
through the Equivalence Theorem, 
\be 
T(\omega^a\omega^b \rightarrow \omega^c\omega^d )  
= T\left(W_L^aW_L^b \rightarrow W_L^cW_L^d\right) + O\left(\frac{M_W}{\sqrt{s}}\right)\ , 
\ee 
and thus $\ar T\ar^2$ is observable.
Therefore we will concentrate on the WBGB scattering for 
$M^2_\varphi, M^2_W, M^2_Z \ll s < \Lambda^2$ where $\Lambda$ is some ultraviolet (UV) cutoff of about  3 TeV, setting the limits of applicability of the effective theory. 
In order to expose the increased strength of the interaction, it suffices  
to treat the scalar dynamics alone, so we will turn off the
transverse electroweak gauge fields by setting $g=g'=0$. 
Then Eq.~(\ref{genericLagrangian}) can be written as 
\begin{eqnarray} \label{bosonLagrangian}
{\cal L} &=& \frac{1}{2}g(\varphi/f) \partial_\mu \omega^a
\partial^\mu \omega^b\left(\delta_{ab}+\frac{\omega^a\omega^b}{v^2-\omega^2}\right)   
\nonumber +\frac{1}{2}\partial_\mu \varphi \partial^\mu \varphi \nonumber \\
&&-\frac{1}{2}M^2_\varphi \varphi^2-\lambda_3 \varphi^3- \lambda_4 \varphi^4+...
\end{eqnarray} 

Notice that by rescaling $f$ and redefining $\beta$ in a trivial way it is possible to set $\alpha=1$ in Eq.~(\ref{gexpansion}) without loosing generality,
\be \label{gexpansion2}
g(\varphi/f)= 1 +2  \frac{\varphi}{f'} +\beta'\left(\frac{\varphi}{f'}\right)^2+\dots
\ee 
in which case the bands on $f/\alpha$ Eq.~(\ref{boundexp1}) and~(\ref{boundexp2}) 
give directly the interval of $f'$ values (around $v$) that are temptatively viable in view of experimental data.
This leaves two free parameters affecting the WBGB's in our energy region of interest (the redefined $f$ and $\beta$), and two more ($\lambda_3$, $\lambda_4$) in the scalar self-potential $V$, assuming that the scalar mass $M_\varphi$ is now as hinted from experiment;  we set $M_\varphi \simeq $ 125 GeV.  However, in the following we will still keep the explicit $\alpha$-dependence in our formulae so that we can easily  trace it for comparison with previous works. In particular the old EWChL without any Higgs-like light resonance corresponds to $\alpha=\beta=0$. 
Nevertheless our numerical results presented in section~\ref{sec:numeric} are all obtained with $\alpha=1$.
Also, as customary in this context, it will be useful to introduce the adimensional parameter $\xi =v^2 /f^2$  so that $\xi=1$ corresponds to the MSM. 

\section{Scattering in perturbation theory} \label{sec:scattering}
\subsection{Feynman amplitudes}

\begin{figure}[h]
\includegraphics[width=14cm]{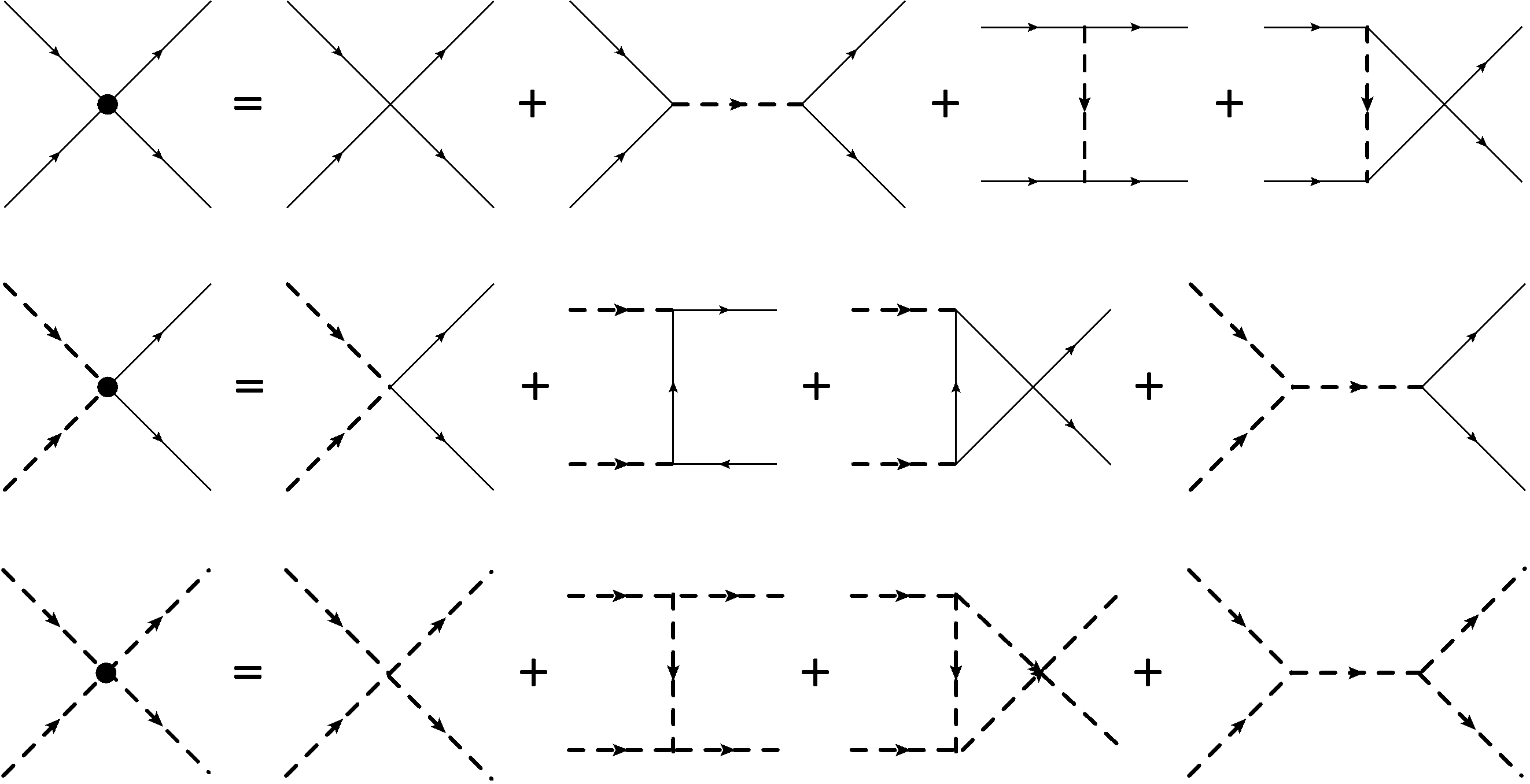}
\caption{\label{fig:Feynmanpert} Feynman diagrams for the tree-level scattering amplitudes of WBGB's (related to $W_L$ by the equivalence theorem) and a light scalar $\varphi$.}
\end{figure}

 With the Lagrangian density in Eq.~(\ref{bosonLagrangian}) it is now straightforward to obtain the tree-level
scattering amplitudes among the four particles that appear at low-energies.
The corresponding Feynman diagrams are represented in Figure~\ref{fig:Feynmanpert}.

First of all we have the WBGB elastic scattering amplitude, usually parametrized in terms of one function $A(s,t,u)$, which in this particular case is a function of
$s$ only: 
\be
T_{abcd}^{2\omega 2\omega}= A(s)\delta_{ab}\delta_{cd}+A(t)\delta_{ac}\delta_{bd}+A(u)\delta_{ad}\delta_{bc}
\ee 
where 
\be A(s)=\frac{s}{v^2}\left(1+ \xi \alpha^2\frac{s}{M_\varphi^2-s}\right) 
\ee
which for the MSM case $\xi=\alpha=1$ leads to the well-known result
\be
A(s)=\frac{s}{v^2}\frac{1}{1-s/M_H^2}\ . 
\ee
In the massless scalar limit  we obtain the low-energy theorem (LET),
\be \label{basicAlinear}
A_0(s)=(1-\xi \alpha^2)\frac{s}{v^2}. 
\ee
which is a generalization, including the new light scalar boson, of the one given in \cite{LET}.  
Interestingly enough, this amplitude vanishes only in the particular
case of the MSM. This fact is an indication of the weakly
interacting nature of the Higgs sector for $M_H \simeq 125 GeV$.
However, in the general case with two scales $v$, $f$, and $\xi \ne 1$,
the amplitude increases linearly with $s$, suggesting the possibility of having a strongly interacting scenario. This observation triggers the rest of our investigation.

Next we consider the inelastic process 
$\omega^a\omega^b \rightarrow \varphi\varphi$ 
or its inverse  $\varphi\varphi \rightarrow \omega^a\omega^b$ 
which have the same amplitude because of the assumed time-reversal invariance of the EWSBS sector.
The tree-level inelastic amplitude is
\ba
T_{ab}^{2\omega 2\varphi}= -\frac{\delta_{ab}}{f^2}
\left[ \beta s
+  \alpha^2\frac{(M_\varphi^2-t)^2}{t}  +\alpha^2  \frac{(M_\varphi^2-u)^2}{u}  +
\frac{6 \alpha \lambda_3  f s }{s-M_\varphi^2}    \right].
\nonumber \\ 
\ea 
In the massless scalar limit
one gets
 \be T_{ab}^{2\omega 2\varphi}=-\frac{\delta_{ab}}{f^2}[ \beta s + \alpha^2 (t+u) + 6 \alpha \lambda_3 f]. 
 \ee 
 By using the on-shell relation $s+t+u=0$ we see that this amplitude is proportional to $\lambda_3$ 
 in the particular case $\alpha^2=\beta$, which includes the MSM and dilaton models.
 However
typically the constants $\lambda_i$ are of the order of $M_\varphi^2$
and then we have $T_{ab}^{2\omega 2\varphi}=0$ in the massless scalar limit. Thus, for dilaton-type models, it is a very good approximation in practice to neglect the channel coupling and concentrate only on the elastic $\omega\omega$ amplitude.
Nevertheless we shall handle the coupling to keep the discussion generic.

The third independent amplitude, scalar-scalar scattering $\varphi\varphi
\rightarrow   \varphi\varphi  $, is again elastic and given by 
\be 
T^{2\varphi 2\varphi}= - 24 \lambda_4-36
\lambda_3^2\left( \frac{1}{s-M_\varphi^2}  +\frac{1}{t-M_\varphi^2}
+\frac{1}{u-M_\varphi^2}    \right) 
\ee
and again we see that $T^{2\varphi 2\varphi}=0$ in the massless scalar limit. 
Therefore in this limit, only the elastic WBGB survives, with amplitude linear in $s$ provided that $\xi$ is different from one, i.e., 
 the MSM excepted. This is clearly an indication for strongly interacting WBGB scattering as a generic property of any model different from the MSM.

For full generality one should notice that  the WBGB interaction is not exactly linear as in Eq.~(\ref{basicAlinear}). 
Following the philosophy of the EWChL it is also possible to add to the original Lagrangian density~(\ref{genericLagrangian}) higher derivative terms. For example one could add the four derivative terms
\be \label{higherorderL}
{\cal L}_4= a_4(Tr V_\mu V_\nu)^2+  a_5(Tr V_\mu V^\mu)^2 
\ee
where $V_\mu= D_\mu U U^\dagger$. These terms produce  additional
contributions to the tree-level EWGB scattering of order $s^2$, and bring-in dependence on two new adimensional parameters $a_4$ and $a_5$ which of
course depend on the underlying dynamics producing the spontaneous electroweak symmetry breaking. 
Following the philosophy of the EWChL these new  terms should be added to the previous leading term at high energies proportional to $s$ in the WBGB scattering amplitude. However a complete consistent treatment requires also the computation of loops which are also of the order of $s^2$ as it was done long time ago in \cite{DHD}. The EWChL parameters $a_4$ and $a_5$ require renormalization which in turn makes them dependent on an arbitrary renormalization scale $\mu$ in such a way that this dependence compensates the one coming from the one-loop 
$\log(-s/\mu^2)$ to make the total WBGB amplitude $\mu$ independent.

\subsection{Partial waves}

The tree-level amplitudes are not appropriate descriptions much above thresholds since they do not respect unitarity. In the following we will introduce and compare different methods for improving the unitary behavior of the amplitudes. Since the unitarity relation is simplest for each partial-wave amplitude, it is convenient to introduce the angular expansion
 \be
 t_{IJ}(s)=\frac{1}{64 \pi}\int _{-1}^1 d(\cos  \theta)P_J(\cos \theta)  T_I(s,\cos \theta)
 \ee
where $ P_J(\cos \theta) $ are the Legendre polynomials, $\theta$ is
the scattering angle in the center of mass frame and $T_I$ are the (custodial)
isospin amplitudes. From now on we will concentrate in the (iso)
scalar channel $I=J=0$. Then for the WBGB scattering we have: 
\be
T_0^{2\omega 2\omega}=  3 A(s)  + A(t) + A(u) 
\ee
 and $P_0=1$. For the process
$\omega^a\omega^b \rightarrow \varphi\varphi$, $T_0^{2\omega 2\varphi}=\sum_a
T_{aa}^{2\omega 2\varphi}/\sqrt{3}$ and for $\varphi\varphi \rightarrow  \varphi\varphi$, obviously $T_0^{2\varphi 2\varphi}  = T^{2\varphi 2\varphi} $. It is simple enough to find the corresponding (iso) scalar amplitudes as 
\begin{eqnarray*}
 t_{2\omega 2\omega}(s) &=& \frac{s}{32\pi v^2}\left(2+\frac{3 \xi \alpha^2 s}{M_\varphi ^2- s}\right)+ \\   
 &&+ \frac{\xi \alpha^2}{32\pi v^2}
\left[s- 2 M_\varphi^2  + \frac{2 M_\varphi ^4}{s}\log\left(1 + \frac{s}{M_\varphi^2}\right)\right],
\label{treetw}
\end{eqnarray*}
\begin{eqnarray*}
 t_{2\omega 2\varphi}(s) &=& t_{2\varphi 2\omega }(s) = -\frac{\sqrt{3}
s}{32\pi f^2}\left(\beta-\frac{6 \alpha\lambda_3  f}{M_\varphi ^2- s}\right) +\\
 &&+ \frac{\sqrt{3}\alpha^2}{32\pi f^2} \left[s + 2 M_\varphi^2   - 
\frac{2 M_\varphi ^4}{s \sigma}\log\left(\frac{2 M_\varphi^2-   
s(1-\sigma)}{2 M_\varphi^2- s (1+\sigma)}\right)\right], 
\end{eqnarray*}
 and 
\begin{eqnarray}
 t_{2\varphi2\varphi}(s) &=& -\frac{1}{16\pi}\left(12
\lambda_4 - \frac{18 \lambda_3^2}{M_\varphi^2-s}\right) -\nonumber\\
 &&- \frac{9 \lambda_3^2}{4 \pi s \sigma^2}\log\left(\frac{2 M_\varphi^2-s(1-\sigma^2)
}{2 M_\varphi^2- s (1+\sigma^2)}\right),
\end{eqnarray}
 where $\sigma = \sqrt{1- 4 M_\varphi^2/s}$.

\subsection{Coupling channels}

The perturbative amplitudes found in the previous subsection have the well known left cuts (LC) coming from the angular integration of the crossed
channels which start in different points for each channel due to the
different thresholds. However, as they are tree-level, real partial
waves in the physical region, they do not carry the full analytical structure; 
they lack an imaginary part for physical $s$, 
and they do not have a right cut (RC) starting at
threshold.  Notice that unitarity is here a non trivial
requirement, coupling the three amplitudes in
a non-linear way.

To simplify the discussion, and also because we are mainly
interested in the region $M_\varphi^2 \ll s$, we will work 
with a slightly simplified version of the partial waves,
corresponding to the asymptotic behavior in the $M_\varphi^2 /s \ll 1 $
regime. This approximation will not only simplify a bit the
equations, but more importantly it will make all of the left cut
thresholds coincide at $s=0$, which will be
important in the following to avoid mixing of LC and RC cuts in
the coupled-channel unitarized amplitudes. Then we will be using  the
approximate tree-level partial waves: 
\ba \label{treewaves}
t_\omega(s)= 
\frac{1}{16\pi v^2}\left[ s(1-\xi \alpha^2)-\xi \alpha^2 M_\varphi^2 +
\xi  \alpha^2 \frac{M_\varphi ^4}{s}\log\left(\frac{s}{M_\varphi^2}\right)\right], 
\ea 
\ba 
t_{\omega\varphi}(s)=t_{\varphi \omega}(s)=
\frac{\sqrt{3}(\alpha^2-\beta)s}{32\pi f^2}+
\frac{\sqrt{3}\alpha^2}{16\pi f^2}\left[M_\varphi^2+ \frac{2 M_\varphi
^4}{s}\log\left(\frac{s}{M_\varphi^2}\right)\right] 
\ea and 
\ba t_{\varphi}(s)=  
\frac{9
\lambda_3^2}{4 \pi s }  \log\left(\frac{s}{ M_\varphi^2}\right)-\frac{3\lambda_4}{4\pi} 
\ea
 where we have used the simplified notation $\omega = 2\omega 2\omega$, $\omega\varphi=2\omega2\varphi$ and
 $2\varphi 2 \varphi= \varphi$.

 The above amplitudes are grouped into the tree-level reaction matrix
 \begin{eqnarray}
T &=&
\left(
\begin{array}{cccc}
 t_\omega & t_{\omega\varphi} \\ t_{\varphi\omega}& t_\varphi
\end{array}\right).
\label{bulkmetric}
\end{eqnarray}

 \subsection{The Electroweak Chiral Lagrangian approach}
\label{subsec:EWChL}
The tree level amplitudes can be approached in a different way by considering the limit
 $M_\varphi^2 \simeq  M_W^2 \simeq M_Z^2 \ll s $ which is appropriate for applying the effective theory ideas. Instead of starting from a non-linear $\sigma$-like model, one can adopt the power-counting of the Electroweak Chiral Lagrangian (EWChL), that we now quickly apply to the case of one additional Higgs-like scalar $\varphi$: In this case the low energy partial waves become
 \begin{eqnarray} \label{LETS}
 t^\omega_0  & = & \frac{s}{16 \pi v^2}(1-\xi \alpha^2)   \nonumber  \\
  t^{\omega\varphi}_0  & =  & \frac{\sqrt{3}(\alpha^2-\beta)s}{32 \pi f^2}   \nonumber  \\
  t^\varphi_0  & = & 0
 \end{eqnarray}
 which are the order $s$ LETs. However, following the philosophy of the EWChL, one should also add order $s^2$ contributions to the WBGB partial wave. To this order  we have tree level contributions coming from the four derivative terms of the effective Lagrangian discussed at the end of  subsection III.A proportional to $a_4$ and $a_5$. Also we have to add the one-loop corrections with WBGB and $\varphi$ internal lines 
which are also order $s^2$. Of course the one-loop integral is divergent 
but the divergences can be regulated, for example by using dimensional regularization, and reabsorbed  by the $a_i$ couplings which become renormalized couplings $a_i(\mu)$ depending on an arbitrary renormalization scale $\mu$. 
 Then the total NLO contribution can be written as:
 \be \label{oneloopChPT}
 t^\omega_1=s^2\left[A(\mu)+ D \log\left(\frac{s}{\mu^2}\right)+ E \log\left(\frac{-s}{\mu^2}\right)\right]
 \ee
where
\be
A(\mu)=A_0+\frac{7 a_4(\mu) + 11 a_5(\mu)}{12 \pi v^4}.
\ee
 Here $A_0$ is a constant depending on the particular regularization scheme used (we will show the details of the computation elsewhere \cite{ours}).  $D$ and $E$ are two constants tuned with $A(\mu)$ so that the amplitude is $\mu$ independent i.e.:
 \be
 A(\mu)=A(\mu_0)+(D+E)\log\frac{\mu^2}{\mu^2_0}
 \ee
  This partial wave is formally similar to the one found in \cite{DHD} but  the constants are different because now 
the light scalar resonance $\varphi$ 
contributes to the sum over intermediate states at one loop. These constants can be obtained from the results found in the very complete recent work \cite{domenec2} where the one-loop scattering amplitudes, including the Higgs-like scalar and WBGB,  are computed.   Using perturbative unitarity we have
 \begin{eqnarray}
 \Imag t^\omega_1 & =  &  \mid t^\omega_0 \mid^2  +  \mid  t^{\omega\varphi}_0   \mid ^2 
  =   \left(\frac{s}{16 \pi v^2}\right)^2\left[(1-\xi \alpha^2)^2 +\frac{3}{4}(\alpha^2-\beta)^2\xi ^2\right] .
 \end{eqnarray}
 On the other hand, from equation~(\ref{oneloopChPT}) above:
 \be
 \Imag t^\omega_1 = - E \pi s^2
 \ee
 so that
 \be
 E= -\frac{\pi}{(4 \pi v)^4}\left[(1-\xi \alpha^2)^2 +\frac{3}{4}(\alpha^2-\beta)^2\xi ^2 \right].
 \ee
Thus the EWChL prediction for the amplitude is:
\be
t^\omega_\chi = t^\omega_0 + t^\omega_1 + \Or(s^3/v^3).
\ee
 This partial wave has better unitary and analytical properties than the tree level amplitude $t^\omega_0$ having the proper LC. However unitarity is only realized perturbatively or in other words:
 \be
 \Imag t^\omega_\chi = \mid  t^\omega_\chi    \mid^2  + \Or(s^3/v^3).
 \ee
We see below how this result from EWChL approach can be enormously improved by the use of the so called Inverse Amplitude Method (IAM).
 
 Another important comment is that in the particular but paradigmatic case of the MSM,  $v=f$ (or $\xi=1$), and the whole
 EWChL approach doesn't make sense reflecting that the MSM with a light Higgs is unitary but weakly interacting at the energy we are interested in here, i. e. $s\gg M_\varphi^2$.

\section{Unitarized scattering amplitudes}\label{sec:unitarity}
The unitarity condition for the exact reaction matrix $\tilde T$ for
massless particles reads 
\be \label{CondicionUnitariedad}
  \Imag \tilde T = \tilde T \tilde T^\dagger
\ee 
on the physical region, i.e., just above the RC where $s= | s | + i
\epsilon$. There are in the literature different ways for
incorporating exact unitarity to a given approximate amplitude and a
lot of controversy about which method is more appropriate or even
the level of arbitrariness of all of them.
To extract model-independent features, we will therefore make use of several of the known methods, that we treat in the following subsections.

\subsection{K-matrix, algebraic unitarization \\
and strength of channel-coupling interaction}\label{subsec:K}

Starting from a tree amplitude, probably the simplest unitarization is the so
called K-matrix method \cite{Gupta}. In our simple case this amount to defining
the unitarized reaction matrix as $\tilde T = T (1- i T )^{-1}$
which obviously satisfies the unitarity condition for real, positive $s$, since $T = T^T= T^{\dagger} $. However this new matrix 
 doesn't  have the proper analytical structure since in particular it lacks a RC. This is a displeasing drawback since physical amplitudes cannot have
poles in the first, or physical, Riemann sheet. The RC opens the door
to analytical prolongation to the second Riemann sheet where poles
belong and may be found and understood as dynamically
generated resonances, so important in strongly interacting scenarios. 
For this reason we will rather define our unitarized matrix, 
(often called K matrix in the literature), as
\be \label{LSalgebraic}
\tilde T = T (1- J T )^{-1}
\ee
 where $J(s)$ is the function
 \be
 J(s)= -\frac{1}{\pi}\log\left(\frac{-s}{\Lambda^2}\right) \ ,
 \ee
and $\Lambda$ some UV cutoff. This function provides the proper analytical structure to $\tilde T$ with the RC  needed for analytical prolongation to the second Riemann sheet. In addition $Im J=1$ on the RC so that the unitarity
   condition is exactly fulfilled. Possible resonances can show up as roots of the determinant $\Delta = \det (1- J T )$,
    or in other words they solve
 \be \label{zeroposition}
 \Delta = 1 - J (t_\omega + t_\varphi )+ J^2 (t_\omega t_\varphi-t_{\omega\varphi}^2)=0
 \ee
 on the second Riemann sheet 
\footnote{The amplitude in Eq.~(\ref{LSalgebraic}) can be thought of as the resummation of the relativistic Lippmann-Schwinger series with the scattering calculated at one loop. Normally this would lead to integral expressions, but the purely algebraic expression in which $T$ and $J$ are factorized (no potentials inside the loop integrals) is a feature of effective theories with polynomial expansions, called on-shell factorization (the off-shell, non-factorizable parts are proportional to tree-level diagrams and can be reabsorbed in the corresponding parameters)~\cite{Oller}.}.

Since the key concern is the WBGB scattering amplitude, 
we write down the unitarity condition for a particular partial wave
\be 
\Imag \tilde t_\omega = \tilde t_\omega \tilde  t_\omega^\ast  + \tilde t_{\omega\varphi}
\tilde t_{\omega\varphi}^\ast. 
\ee This equation is exactly  fulfilled by the
corresponding given angular-momentum matrix element of reaction matrix
$\tilde T$ of Eq.~(\ref{LSalgebraic}),
\be  \label{improvedK}
\tilde t_\omega = \frac{t_\omega -
J (t_\omega t_\varphi-t_{\omega\varphi}^2)}{1 - J (t_\omega + t_\varphi )+
J^2 (t_\omega t_\varphi-t_{ \omega\varphi}^2)}. 
\ee
In Figure~\ref{smallCoupled}  in section~\ref{sec:numeric} below we show that the effect of the coupled $\varphi\varphi$ channel on the $W_LW_L$ channel is small, in the particular case $\beta=1$, as can also be seen analytically.
As we have already pointed out, the partial waves $t_{\omega\varphi}$ and $t_\varphi$ are then suppressed when $M_\varphi^2  \ll s$,
the energy region of interest in this work. That means that
 $t_{\omega\varphi}$ and $t_\varphi$ are subleading with respect to $t_\omega$ and therefore one could  expect to have:
 \be
 \Imag \tilde t_\omega \simeq \tilde t_\omega \tilde  t_\omega^\ast
 \ee
 on the RC and this is indeed the case, as also checked numerically in Figure~\ref{smallCoupled}.

From Eq.~(\ref{LETS}) it is clear that for $\beta=\alpha^2$ the WBGB are
 very weakly coupled to the scalars  $\varphi\varphi$ and one can consider the WBGB elastic scattering alone. However, in the more general situation $\beta \ne \alpha^2$ the WBGB can be strongly coupled 
 to the $\varphi\varphi$ state and then a coupled channel formalism is needed to study the $\omega\omega$ interactions. For the particular but interesting case $\beta=\alpha^2$ (elastic case) the K-matrix unitarized partial wave $t_\omega$
reduces to the simple algebraic formula
 \be\label{unitscattering}
\tilde t_\omega=\frac{t_\omega}{1- J t_\omega}.
 \ee
 Now we can again take $M_\varphi \simeq 125$ GeV, leaving only as free parameter the new scale $f$. Notice that this case includes in particular all  dilaton models.

We have now sufficient theory to address the interesting question 
of when are the interactions weak or strong, and the question of whether any poles appear in the second Riemann sheet of the amplitude.

Advancing generic results of section~\ref{sec:numeric}, we can see for example in Figures~\ref{fig:unitscattering2} and
~\ref{fig:polef400}
that for any value of $f$ not too close to $v$, we are typically in a strongly interacting regime (with partial wave with modulus of order one) and also produce a pole in the second Riemann sheet. In QCD, setting $\xi =0$
and $v=f_\pi$  in the low energy theorem for $A_0(s)$ above, we would be reproducing a K-matrix unitarized first order ChPT for pions in the chiral limit and then the found resonance would be just the $\sigma$-meson \cite{donoghue}. 

However, as we can see in Fig~\ref{fig:polemotion}, the width corresponding to the  pole found is in general too large to be considered a real resonance, but still produces such huge enhancement of the scattering, that it leads to a strong-interaction regime for most conceivable $f$-values. 
The exception is the region $f \simeq v$. This precisely  corresponds to the renormalizable MSM. Of course the MSM with a light Higgs of 125 GeV is obviously weakly interacting.

The surprise is that in the larger class of models considered in this work,
that one needs to examine following the effective-theory philosophy, 
 this weakly interacting behavior is quite rare since a kind of fine-tuning $f=v$ is needed to avoid the generic  strongly interacting scenario.

 \subsection{Alternative unitarization: large N limit}\label{subsec:largeN}

Eq.~(\ref{LSalgebraic}), the core of the K-matrix method, is in the end a model. Although it incorporates the correct low-energy amplitude and is by construction unitary and has the correct RC, it entails an on-shell factorization 
($JT$ is an algebraic matrix product and not an integral operator product) that is strictly true only for polynomial interactions, and the LC is only treated  perturbatively. These are very sensible approximations that capture the main physics for physical $s$ and have been very successful in many cases. Still it is very important to explore  other ways to unitarize the high-energy WBGB scattering amplitude, to discard artifacts of the unitarization method and give more robustness to our general  results. 

In this subsection we rederive an interesting alternative for the computation of the amplitudes, treating them in the large-N limit (where $N$ refers to the number of Goldstone bosons). This is a non perturbative approximation which was introduced for the Linear Sigma Model (LSM)~\cite{Dobado:1996pp}  but it can also be extended to the NLSM~\cite{Dobado:1992jg}. Being non perturbative, the large-N approximation produces, even at lowest order, amplitudes which have good unitarity and analytical behavior so that they don't need any further unitarization. 
The main idea is to extend the original minimal coset for a weak isospin triplet plus a scalar singlet, $SO(4)/SO(3)=S^3$, to  $SO(N+1)/SO(N)=S^N$ so that we now have $N$ WBGB and one scalar. Now one can consider the limit $N\to \infty$ simultaneously with  $v\to 0$ keeping $v^2/N$ fixed (and at the end evaluate the leading formulae thus obtained at $N=3$). 

\begin{figure}[h]
\includegraphics[width=16cm]{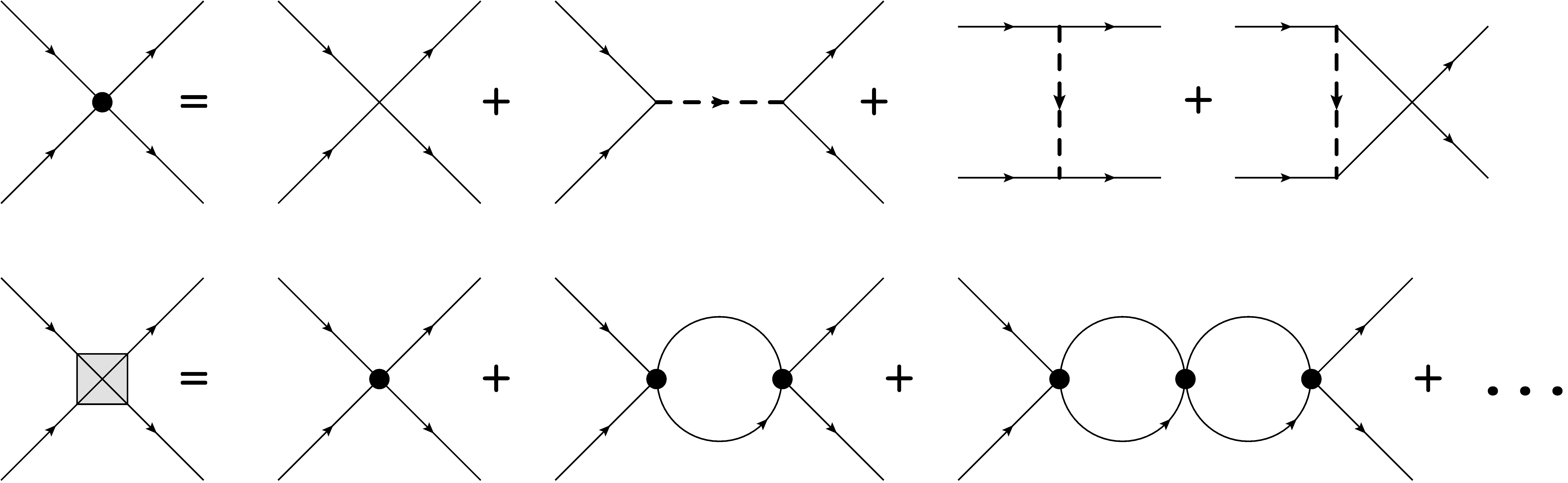}
\caption{\label{fig:FeynmanN} Feynman diagrams for large $N$ scattering WBGB's (related to $W_L$ by the equivalence theorem) and a light scalar $\varphi$.}
\end{figure}

One proceeds by expanding the amplitudes in powers of $1/N$, and in the simplest approach one can keep just the first non-trivial term. Starting from the tree level amplitude for WBGB elastic scattering $A(s)$, and taking into account that $v^2$ is of order $N$, the 
amplitude $A_N$, correct to order $1/N$, is given by a Lippmann-Schwinger series
\be
A_N= A - A \frac{NI}{2} A + A \frac{NI}{2} A\frac{NI}{2} A- A\frac{NI}{2} A\frac{NI}{2} A\frac{NI}{2} A+ \dots
\ee
 where $I=I(s)$ is the simple two body  loop integral:
 \ba
 I(s)=  \int \frac {d^4  q}{(2 \pi)^4}\frac{i}{q^2(q+p)^2} 
= \frac{1}{16 \pi^2}\log\left(\frac{-s}{ \Lambda^2}\right) = 
-\frac{1}{8\pi} J(s) 
 \ea
 and where $p$ is the total momentum so that $p^2=s$; we have introduced the UV cutoff $\Lambda$; $N$ counts the number of WBGB in the loop; and the factors $1/2$
are combinatorial factors because of the indistinguishability of the WBGB. It is immediate to check that all the terms in the series are of order $1/N$. Now it is possible to formally sum the geometric series yielding
\be
A_N(s)=\frac{A(s)}{1 + \frac{N I(s)A(s)}{2}}.
\ee
The corresponding Feynman diagrams are represented in Figure~\ref{fig:FeynmanN}.

The isospin scalar amplitude is given in this case by:
\be
T_0^{\omega}= N A(s) + A(t) + A(u)
\ee
so in the large N limit only the $s$ channel contributes. Thus the (iso)scalar partial wave for elastic WBGB scattering, $t_N^{w}$,  is just
\be \label{fullLargeN}
t_N^{\omega}(s)=\frac{t_0^\omega}{1 - J t_0^\omega}
\ee
where
\be \label{treelargeN}
t_0^\omega(s) =\frac{s N}{32 \pi v^2}\left(1+ \xi \alpha^2\frac{s}{M_\varphi^2-s}\right)
\ee

It is amazing to realize how similar this partial wave and the one obtained from the K-matrix method are. 
 Even if they are not identical for $N=3$, as seen comparing Eq.~(\ref{treewaves}) with Eq.~(\ref{treelargeN}), the geometric-series structure of equations
(\ref{unitscattering}) and (\ref{fullLargeN}), coming from the right $s$-channel cut dominating over the $t$-channel cut for physical $s$ (much closer to the RC), is the same. Thus they give rise to qualitatively and quantitatively similar results for $s\gg M_\varphi^2$ as can be seen in 
Figure~\ref{fig:largeNcomp}.

We thus find that the large-N approximation provides
strong support to the scenario of strongly interacting scattering for the WBGB that we obtain, under the general condition $f\ne v$, from the simple K-matrix amplitude in subsection~\ref{subsec:K}. 

Finally it is interesting to notice that in the large-N limit, the decoupling of the $\omega\omega$ and $\varphi\varphi$ channels occurs in a natural way, because the $\varphi$ loops are suppressed by a factor $N$ with respect to the WBGB loops. So here, the values of $\beta$ and $\lambda_i$ in the $V(\varphi)$ potential are irrelevant to the strength of the $t^{\omega\varphi}$ interaction, that is weak due to the large-N approximation.

\subsection{Alternative unitarization: the N/D method:
\\ algebraic treatment} \label{subsec:NoverD}

To further check that indeed our results are generic, 
and not contingent on the algebraic K-matrix formalism,  
we will also use the N/D method for unitarizing the elastic 
$t^\omega$ scalar partial wave, directly in the elastic case
(i.e. we decouple the $\varphi\varphi$ channel by restricting ourselves to $\beta=\alpha^2$).

We will try to unitarize the perturbative tree-level elastic   
$\omega\omega$ partial wave,  $t^\omega(s)$, calculated in section \ref{sec:scattering}. 
The full $\tilde t^\omega$ amplitude satisfies a dispersion relation given its imaginary parts on the LC, RC, as well as the contribution of any CDD poles or subtraction constants that may be necessary. 

The N/D method is a (non-unique) construction that solves the dispersion relation satisfied by the partial-wave amplitude $\tilde t^\omega(s)$ starting with an ansatz 
as~\cite{Chew}
\be
\tilde t^\omega(s)=\frac{N(s)}{D(s)}
\ee  
where the numerator function
$N(s)$ carries the amplitude's left hand cut, so that $\Imag N(s>0) =0$, and the denominator function $D(s)$ carries the right hand one, and $\Imag D(s<0)=0$.
This gives the quotient $\tilde t^\omega(s)$ the expected analytical structure.  Therefore $\Imag N(s)=0$ on the RC and $\Imag D(s)=0$ on the LC. In addition elastic unitarity requires $\Imag D(s) = - N(s)$ on the RC and also we have $\Imag N(s) = D(s) \Imag\tilde t^\omega(s)$ on the LC. 

The dispersive representation of $\tilde t^\omega$ usually entangles the left and right cuts (as known since the Kramers-Kronig equations in electrodynamics) but the $N/D$ ansatz simplifies the system
 to two equations that are sequentially solved, one to be satisfied by  
$N(s)$, and one yielding $D(s)$ once $N(s)$ has been calculated. 

In once-subtracted form, using the normalization $D(0)=1$, we have
\ba \label{NoverD}
D(s) & = & 1- \frac{s}{\pi}\int_0^\infty  \frac{ds' N(s')}{s'(s'-s-i\epsilon)} \label{solveD}\\
N(s) & = & \frac{s}{\pi}\int_{-\infty}^{0}  \frac{ds' \Imag N(s')}{s'(s'-s-i\epsilon)}  \label{solveN} \ 
\ea
where we  have set to zero the subtraction constant in the equation for $N(s)$  ($N(0)=0$).
Once solved, the complete amplitude is constructed as $\tilde t^\omega(s)=N(s)/D(s)$. As usual with dispersive approaches, these two equations encode the requirement of analyticity (hence causality) but they don't determine the amplitude since they have many solutions. However, starting from some approximated $N(s)$ we can iterate these
equations to find an improved amplitude with  better analytical and unitary properties. For example we can start from some  approximate $N_0(s)$ 
and using  Eq.~(\ref{solveD}) we obtain $D_0(s)$. Then we can use  Eq.~(\ref{solveN}) to get $N_1(s)$ and so on. This  iteration is known to converge very quickly~\cite{Hikasa:1993cg}
in many cases so we will just consider here the lowest order approximation $t^\omega(s) \simeq N_0(s)/D_0(s)$ with $N_0(s)$ being the tree level amplitude.
As the integrals diverge at large and small $s$ and we employ
an IR cutoff $m^2$ and an UV cutoff $\Lambda ^2$ in lieu of a more sophisticated subtraction scheme.

Quite remarkably, it is possible to carry out all integrations of this first iteration analytically for the scalar channel that we are treating, as we now show, delaying some computational details to~\ref{app:Integrals}.

We find 
\be
D_0(s)=1-\frac{s}{\pi}\int _{m^2}^{\Lambda^2}\frac{d s'
t_\omega(s')}{s'(s'-s-i \epsilon)}=1-\frac{M_\varphi^2}{32 \pi^2 v^2}x
I(x) 
\ee 
where $x=s/M_\varphi^2$ and we have introduced the dispersive integral 
\be I(x)=\int_{x_1}^{x_2}\frac{dx'}{x'-x-i\epsilon}\left[2+\xi
-\frac{2\xi}{x'}+ \frac{3 \xi
x'}{1-x'}+\frac{2\xi}{x'^2}\log(1+x')\right] 
\ee 
with the dimensionless $x_1= m^2/M_\varphi^2$ and $x_2=\Lambda^2/M_\varphi^2$ IR and UV regulators. 
This integral can be split in four pieces $I=I_1+I_2+I_3+I_4$ for ease of handling.
The first is 
\be I_1= (2+\xi)\log\frac{x_2-x}{x_1-x}\ ;
\ee 
Notice that, because of $x_1\le x \le x_2$, $I_1$ develops an imaginary part on the RC. The same happens to the following integrals too. 
\be
I_2  =  \frac{2\xi}{x}\log\frac{x_2(x_1-x)}{x_1(x_2-x)}\ , 
\ee 
\be
  I_3  =\frac{3\xi}{1-x}
\left(x\log\frac{x_2-x}{x_1-x}-\log\frac{1-x_2}{1-x_1}\right) 
\ee 
and finally
\begin{eqnarray}
   I_4 &=&\frac{2\xi}{x}\left\{ \frac{\log(1+x_2)}{x_2}- \frac{\log(1+x_1)}{x_1}-\log\frac{x_2}{1+x_2}+\log\frac{x_1}{1+x_1} + \right.\nonumber\\
&&+\left. \frac{1}{x}[Li_2(-x_2)-Li_2(-x_1)] + \frac{1}{x}\left[ \log(1+x_2)\log\frac{x-x_2}{1+x}- \right.\right.\nonumber\\
&&-\left.\left. \log(1+x_1)\log\frac{x-x_1}{1+x} + Li_2\left(\frac{1+x_2}{1+x}\right) -Li_2\left(\frac{1+x_1}{1+x}\right) \right] \right\}
\end{eqnarray}
where we have assumed $x_1>1$ to avoid the zeroes of the  $x-1$ denominators. From the $I_1\dots I_4$ integrals
 it is not difficult to show explicitly that, over the RC,
\be
\Imag D_0 = -N_0 =- t^\omega\ ,
\ee
and therefore, since $\tilde t^\omega = D_0/N_0$ we see that one-channel unitarity over the physical RC is respected,
\be \label{unitarityinNoverD}
\Imag \tilde t^\omega = \mid \tilde t^\omega   \mid  ^2 \ .
\ee
 Thus, $\tilde t^\omega$ is an unitary partial wave with the appropriate analytical structure including its LC and the most important RC. It is then possible to continue analytically to the second Riemann sheet where dynamical resonances could show up as poles at some $s_0=M^2-i \Gamma M$ (with mass and width interpretation reliable as long as $s_0$ is not very far from the real axis, so that $\Gamma  \ll M$).

\subsection{The Inverse Amplitude Method}

In section~\ref{subsec:EWChL} we saw that the EWChL approach provides the first terms of a power series expansion for the WBGB scattering amplitude in  $s/v^2$. The first two terms construct a partial wave which has good analytical properties but still  is unitary only in the perturbative sense.  However the truncated series can be regarded in a much more efficient way by using the so called inverse amplitude method (IAM). This method was introduced in the context of ordinary ChPT for pions 
\cite{DHT} providing a fully unitarized amplitude with the right analytical structure that can be prolongated to the second Riemann sheet and is therefore able to reproduce dynamical resonances as poles in this second sheet. The  use of this method has made possible to fit an enormous set of meson scattering data, including resonances, in terms of a few chiral parameters. Because the EWChL has the same formal structure as the ordinary ChPT one, the application of this method can be extended to this case too \cite{DHT2} (see also \cite{libronaranja} for a complete review of both applications of the IAM). It is based on the dispersion relation fulfilled by the algebraic inverse of the partial waves resulting in exact treatment of the RC contribution and a perturbative estimation of the LC one. After some quite general assumptions concerning the absence  of zeros in the original amplitudes~\footnote{The error incurred by ignoring the subthreshold Adler-zeroes in the simple formula in Eq.~(\ref{IAMeq}) has been corrected for by calculating without this simplification~\cite{GomezNicola:2007qj}, and found to be below the permille level for physical $s$.}, the final result is a very practical expression:
\be \label{IAMeq}
\tilde t^\omega = \frac{t_0^\omega}{1-\frac{t_0^\omega}{t_1^\omega} } .
\ee
As already mentioned, this amplitude is exactly unitary and it has the proper analytical structure including the LC and the RC with $\Imag\tilde t ^\omega= \mid  \tilde t^\omega   \mid ^2$ on it. The analytical  continuation to the second sheet may or may not have poles in the different channels (putative resonances) depending on the values of the renormalized chiral parameters $a_4$ and $a_5$, as studied in \cite{DHT2} for the case of no light Higgs-like resonance and more recently in \cite{domenec1} for the case of the MSM with a light Higgs.

From the very definition of the IAM method and the form of EWChL amplitudes, the position of the poles in the second Riemann sheet $s_0$ fulfill the equation
\be
\frac{16 \pi v^2 s_0}{1-\xi \alpha^2 }(  A(s_0) + i E( Arg(s_0)-\pi   ) )  = 1
\ee
where we have chosen the renormalization scale $\mu$ as the pole position $s_0$ for the sake of simplicity, using the fact that $s_0$, as any observable, is $\mu$ independent. Since $\alpha=1$, for $\xi\to 1$ ($f\to v$) the pole that produces strong interactions is removed to infinite mass, returning the weakly-interacting MSM.

\section{Numerical results} \label{sec:numeric}

In this section we offer several typical numerical results obtained with the different unitarization methods that we have introduced. As we will see, even if some numerical details are different, the qualitative behavior of the amplitudes and their analytical continuation to the second Riemann sheet of $s$ are similar.

\subsection{Improved K matrix approach}

\begin{figure}
\centerline{\includegraphics[width=7cm]{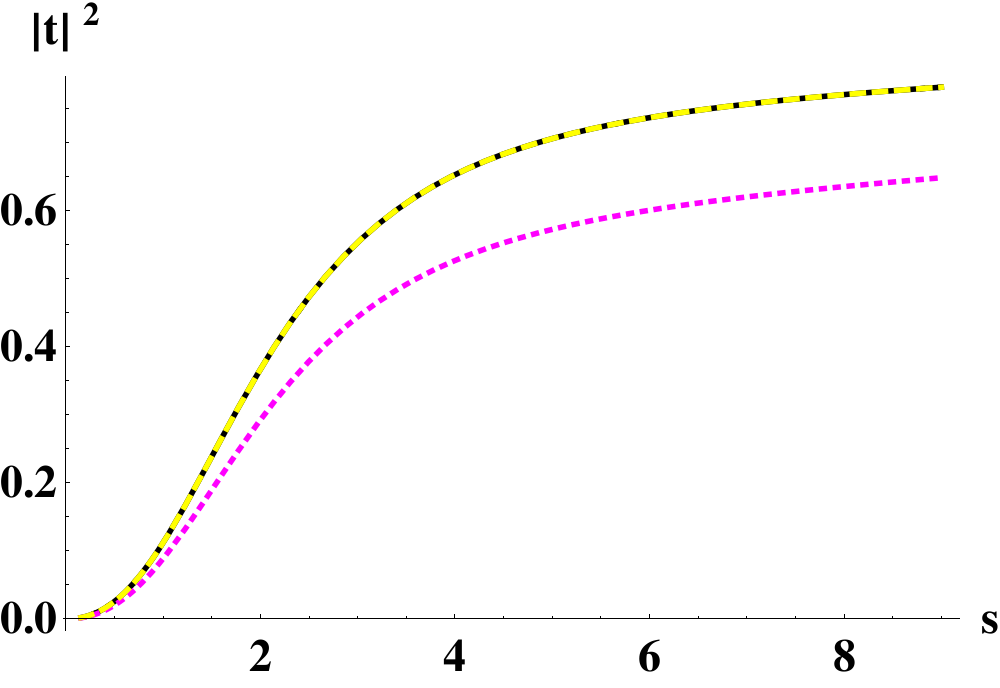}}
\caption{\label{fig:strongchannelcoupling}
Square modulus of the unitarized $t_\omega$ (bottom, dashed line) that fails to satisfy single-channel unitarity for $\alpha=1$, $\beta=3$. 
On the contrary, the sum of the squared scattering s-wave amplitude over the two open channels $\ar t_{\omega}\ar^2 + \ar t_{\omega\varphi}\ar^2$ falls on top of $\Imag t_\omega$ (top two lines, solid and dotted), showing how coupling channels is needed to restore unitarity. See the particular values of the different parameters considered in the main text.
}
\end{figure}

In Figure~\ref{fig:strongchannelcoupling} we show the unitarized $\ar \tilde t_w\ar$ (the WBGB scattering amplitude) that fails to satisfy unitarity unless the second channel is included (also shown in the Figure for a generic $\alpha=1$, $\beta\ne 1$ taken as $\beta=3$, with $\xi\simeq 1/4$, $\lambda_3=M_\varphi^2/f$, and $\lambda_4=M_\varphi^2/f^2$). This is due to the strong interchannel coupling. In the same Figure~\ref{fig:strongchannelcoupling} we plot, in addition to  $\ar \tilde t_{\omega}\ar^2$ (bottom line), $\ar \tilde t_{\omega}\ar^2 + \ar \tilde t_{\omega\varphi}\ar^2$ and $\Imag (\tilde t^\omega)$ (top two lines). The latter two are on top of each other, establishing coupled-channel unitarity.  Notice that all the interactions are strong in this generic case.

\begin{figure}
\centerline{\includegraphics[width=8.5cm]{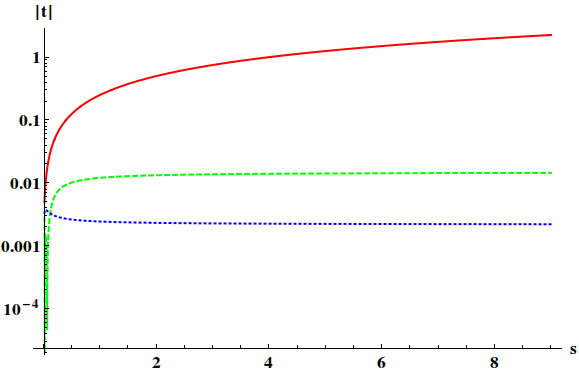}}
\caption{\label{smallCoupled} 
We show the small effect of the coupled channels with the tree-level amplitudes in Eq.~(\ref{treewaves}) for moderate $\beta$ of order $\alpha^2=1$. As seen, the amplitude for $\omega\omega$ scattering (solid line at the top, red online) is much larger than that for the  $\varphi\varphi$ channel (dashed line in the middle, green online) and the interchannel coupling amplitude is smallest (dotted line at the bottom, blue online). Parameters used: $f=2v$, $\beta=\alpha^2=1$, $\lambda_3=M^2_\varphi/f$, $\lambda_4=M^2_\varphi/f^2$. The $OX$ axis represents Mandelstam $s$ in TeV$^2$.
}
\end{figure}

The situation is simpler for $\beta=\alpha^2$ (=1) because the tree-level amplitude $t_{\omega\varphi}$ fails to have a term proportional to $s$ 
and then it is not strongly interacting anymore.  As shown in Figure~\ref{smallCoupled},  for $f\ne v$ the modulus of the elastic amplitude $|t_{\omega}|$ is large and of order 1 for TeV energies. The scalar-scalar interaction remains perturbative since it is controlled by $\lambda_3$ and $\lambda_4$ which are typically of order $M^2_\varphi/f$ and $M^2_\varphi/f^2$ that, for $f>v$, is even weaker than in the standard model case. Finally, the smallest amplitude corresponds to the interchannel coupling $t_{\omega\varphi}$. Thus, for values of $\beta$ near $\alpha^2=1$ it is a good approximation to neglect the coupled channels and use  the one-channel equation~(\ref{unitscattering}) instead of Eq.~(\ref{improvedK}).

Notice that even if we can neglect interchannel coupling, whenever $\beta = \alpha^2$ (or simply $\beta=1$ since, as mentioned above, we can always redefine $f$ and $\beta$ so that $\alpha=1$), 
we will find strong interaction and the need for unitarizing the $\omega\omega$ elastic scattering amplitude.  
Our results are then in broad agreement with other typical works~\cite{Alboteanu:2008my} that considered $W_L W_L$ elastic scattering and also found heavier resonances due to the strong interaction developing.

\begin{figure}
\centerline{\includegraphics[width=7cm]{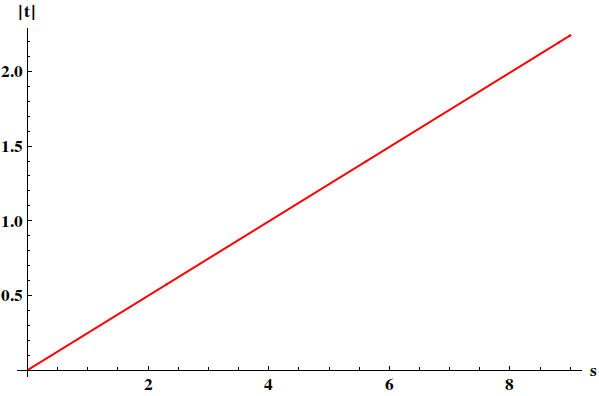}}
\caption{\label{fig:scattering1}
Modulus of the tree-level $t_\omega$ from Eq.~(\ref{treetw}) for physical $s$ showing the violation of unitarity for $f\ne v$. Axes and parameters as in Figure.~\ref{smallCoupled}.}
\end{figure}

As seen in Figure~\ref{fig:scattering1}, unitarity is violated by $t_{\omega}$ whose modulus  exceeds 1 slightly before $\sqrt{s}=2$ TeV for the given  parameter set.
Since this scattering amplitude between WBGB is thus found to be strong, the tree-level approximation  breaks down and some unitarization method is required to have a realistic amplitude usable in comparisons with future experimental data in $\omega\omega$ elastic scattering.

\begin{figure}
\centerline{\includegraphics[width=7cm]{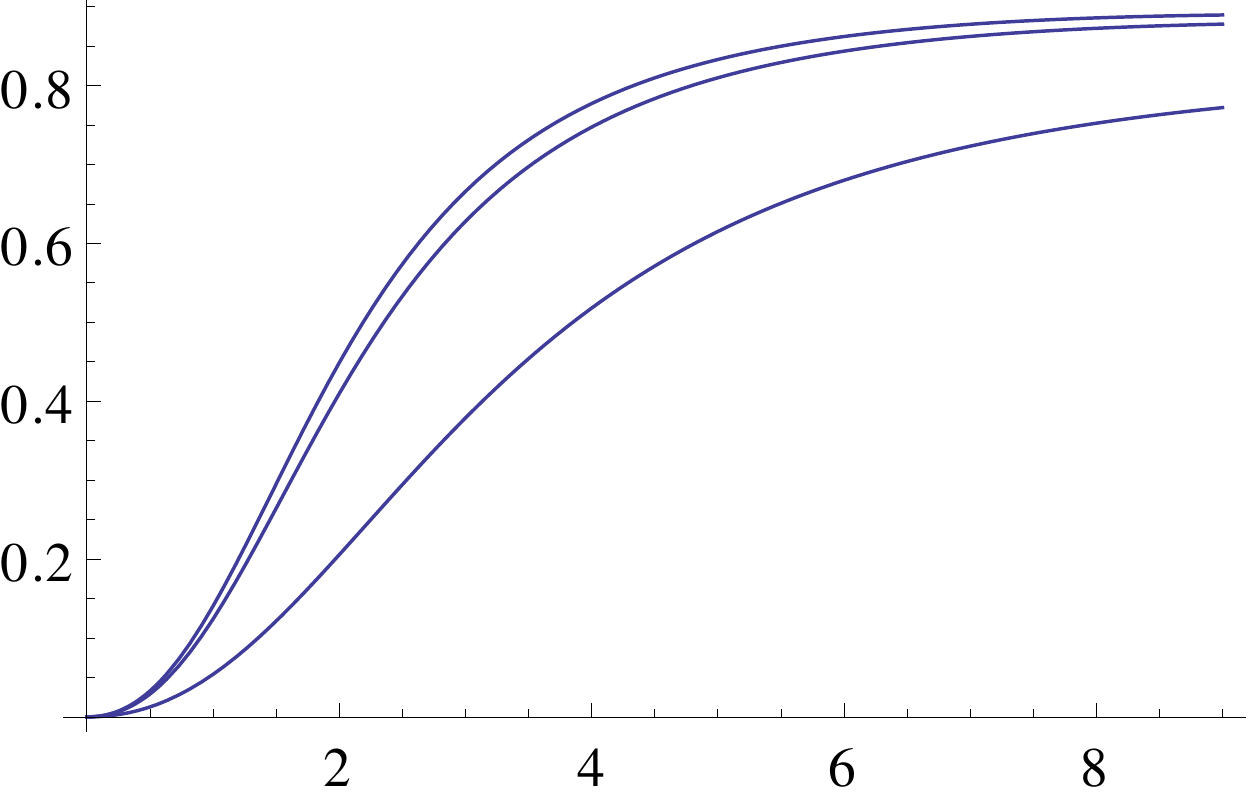}}
\caption{\label{fig:unitscattering2}
Modulus of the unitarized $t_\omega$ from Eq.~(\ref{unitscattering}) for physical $s$ showing that, for $f\ne v$, the appropriately unitarized interactions are strong (with modulus of order 1).
We swipe several values of $f$. From top to bottom, $f=1.2$ TeV, $f=0.8$ TeV, and $f=0.4$ TeV (with $\Lambda=3$ TeV, $\mu=100$ GeV regulating the loop integral).
}
\end{figure}

We therefore unitarize the amplitude, as in Eq.~(\ref{unitscattering}), and plot it in Figure~\ref{fig:unitscattering2}, that is acceptable from the point of view of unitarity, though the interaction is strong (with $|t_\omega|=\Or(1)$).

The amplitude modulus in Figure~\ref{fig:unitscattering2} is indeed characteristic of strong interactions. We have numerically checked that the outcome of the program satisfies $\Imag t = |t|^2$ for null interchannel coupling, and  Eq.~(\ref{CondicionUnitariedad}) for finite interchannel coupling.

In figure~\ref{fig:unitscattering2} we see that the dependence in $f$ is not very pronounced unless $f\simeq v$ (we will analyze this dependence in more detail in subsection \ref{subsec:numericND} below).

\begin{figure}
\includegraphics[width=7cm]{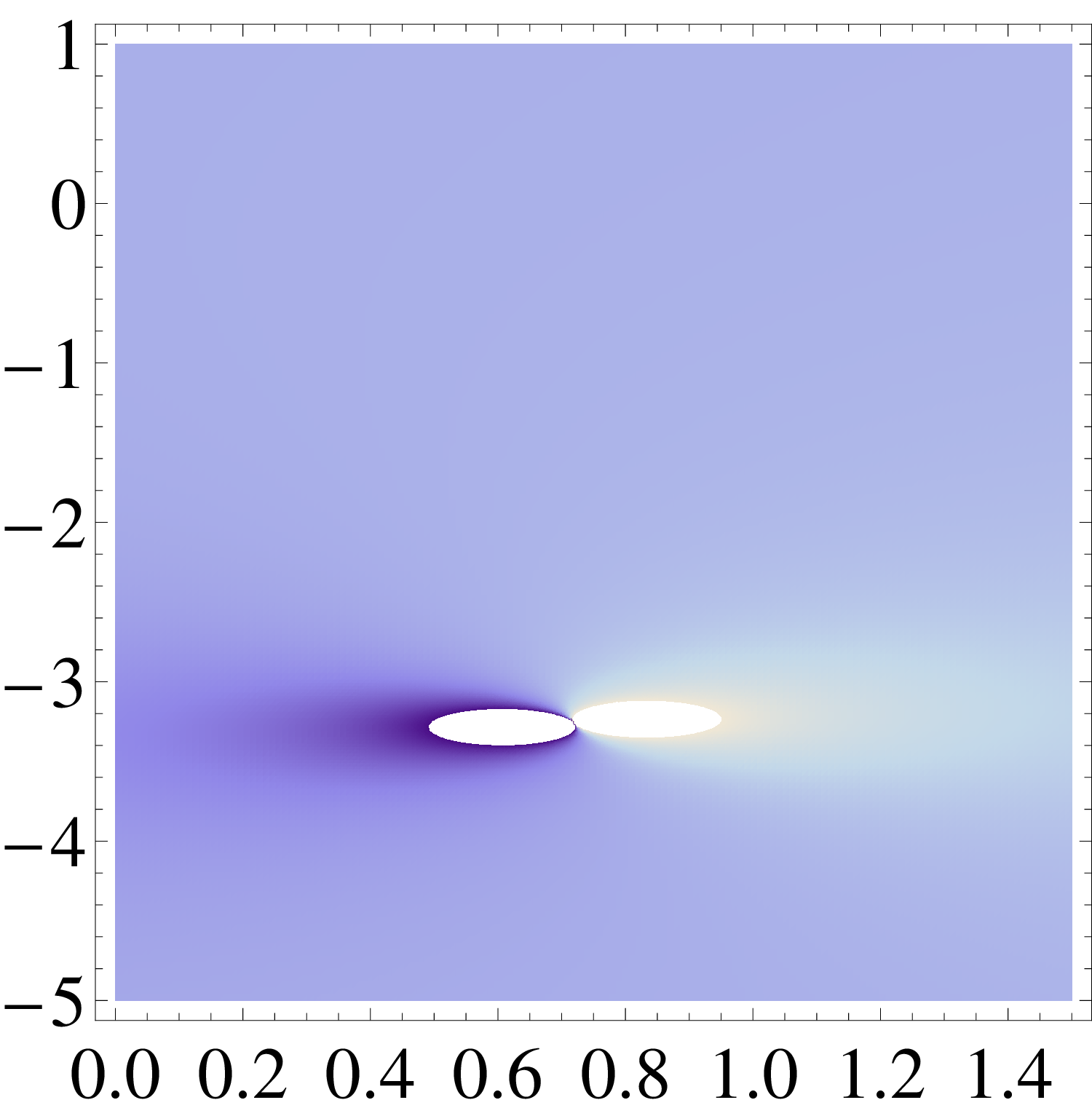}
\includegraphics[width=7cm]{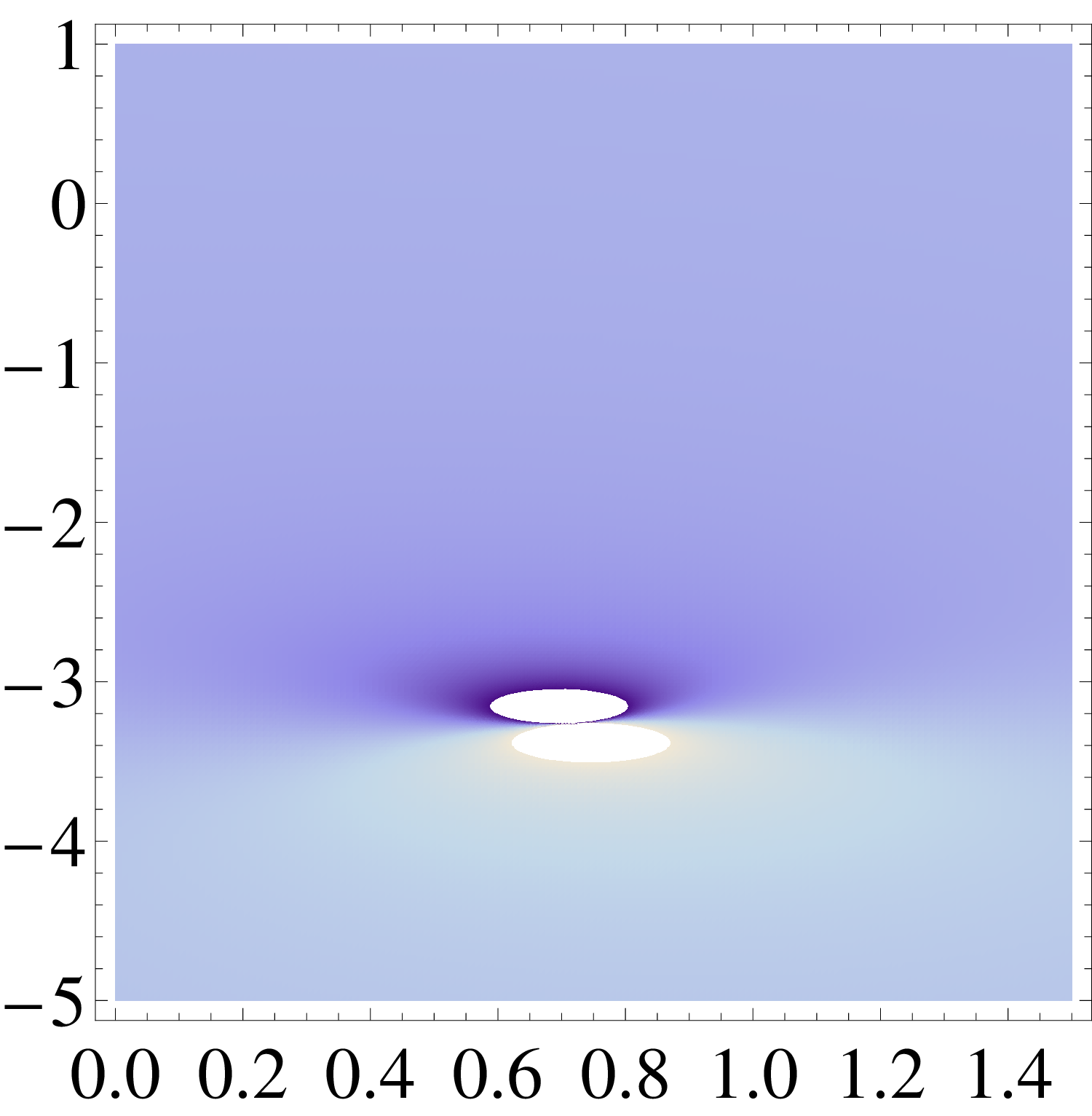}\\
\includegraphics[width=8.5cm]{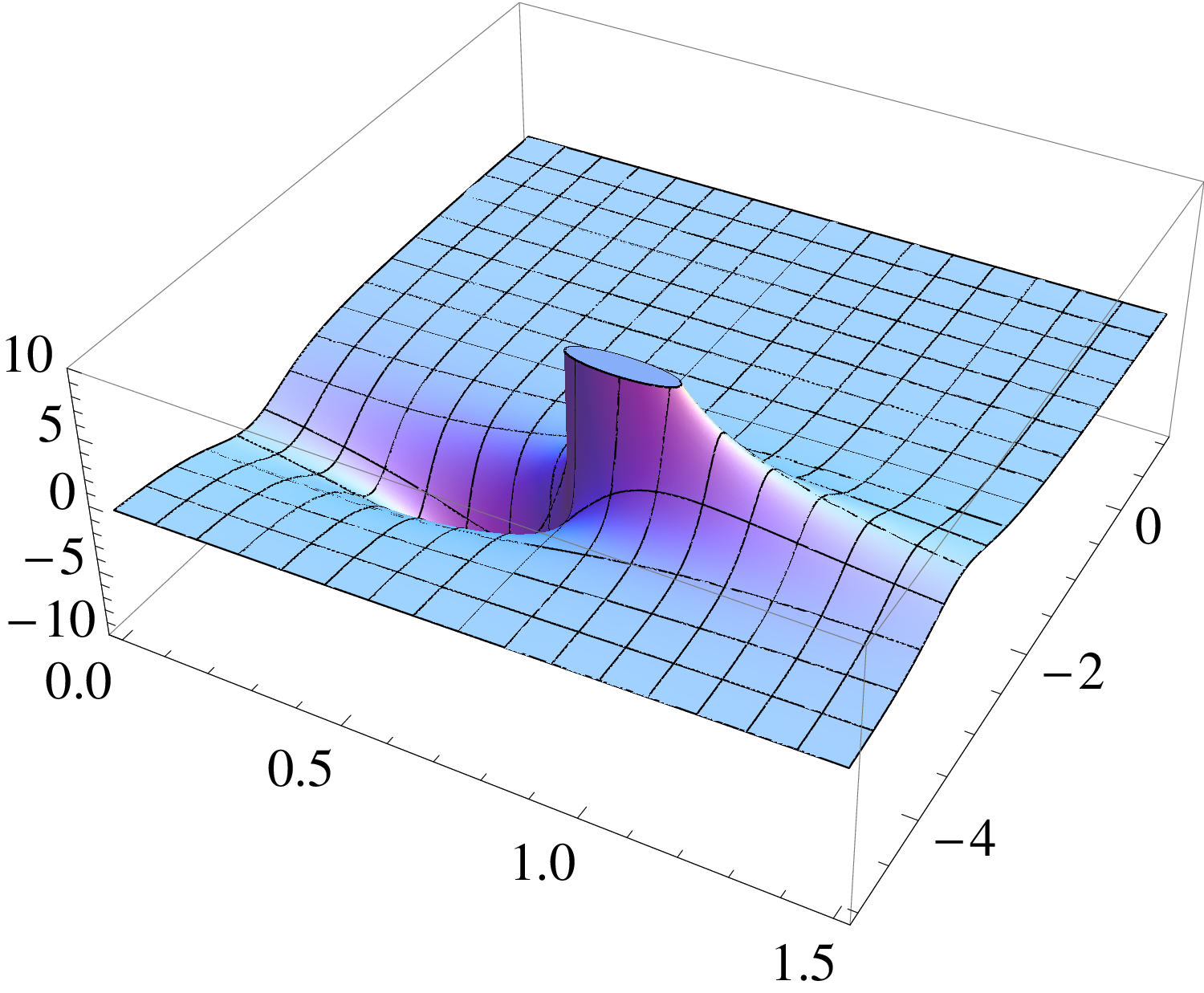}
\includegraphics[width=8.5cm]{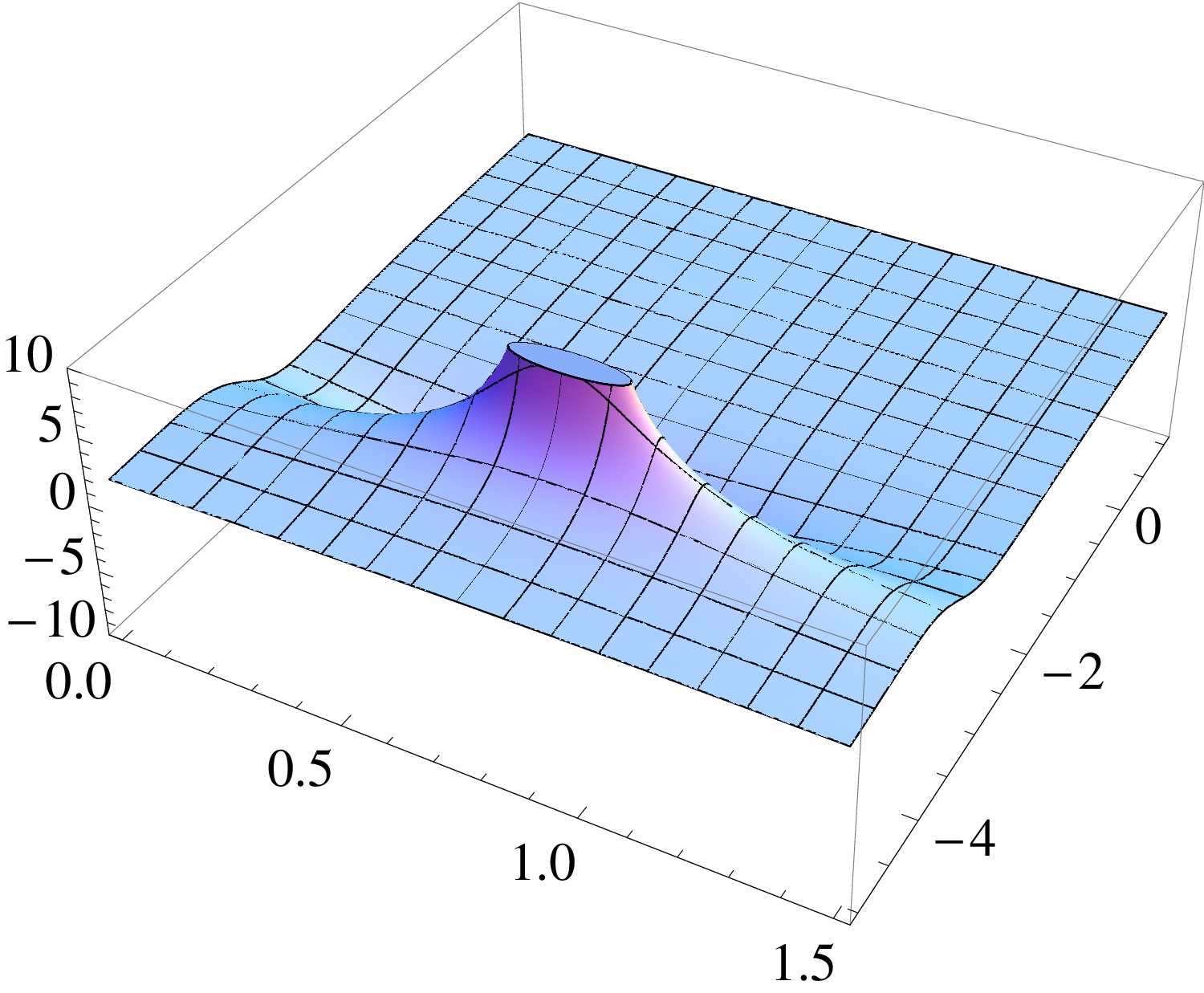}
\caption{\label{fig:polef400} By extending $s$ to the second Riemann sheet, we find that the amplitude presents a pole very far off from the physical axis, reminiscent of the $\sigma$ meson of $\pi\pi$ scattering. Top: two-dimensional contour plots in the complex $s$-plane. Bottom: amplitude against complex-$s$ (real part starting at 0, imaginary part on the second Riemann sheet for negative values, axis units in TeV). The left plots show the real part of the amplitude $Re t^\omega(s)$ and the right plots the imaginary part $\Imag t^\omega(s)$. Here we take $f=400$ GeV not that much larger than $v$.
}
\end{figure}

\begin{figure}
\includegraphics[width=7cm]{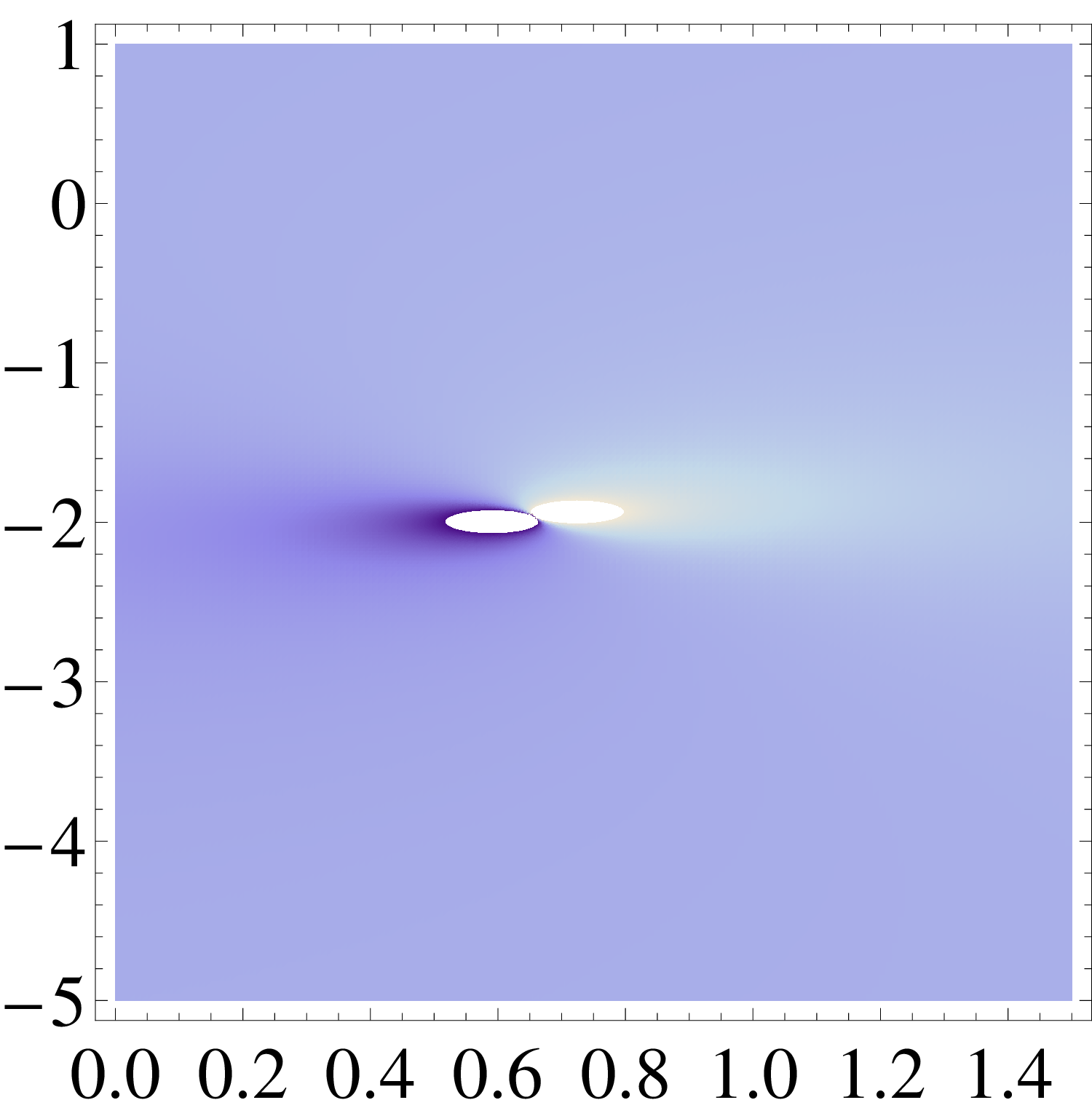}
\includegraphics[width=7cm]{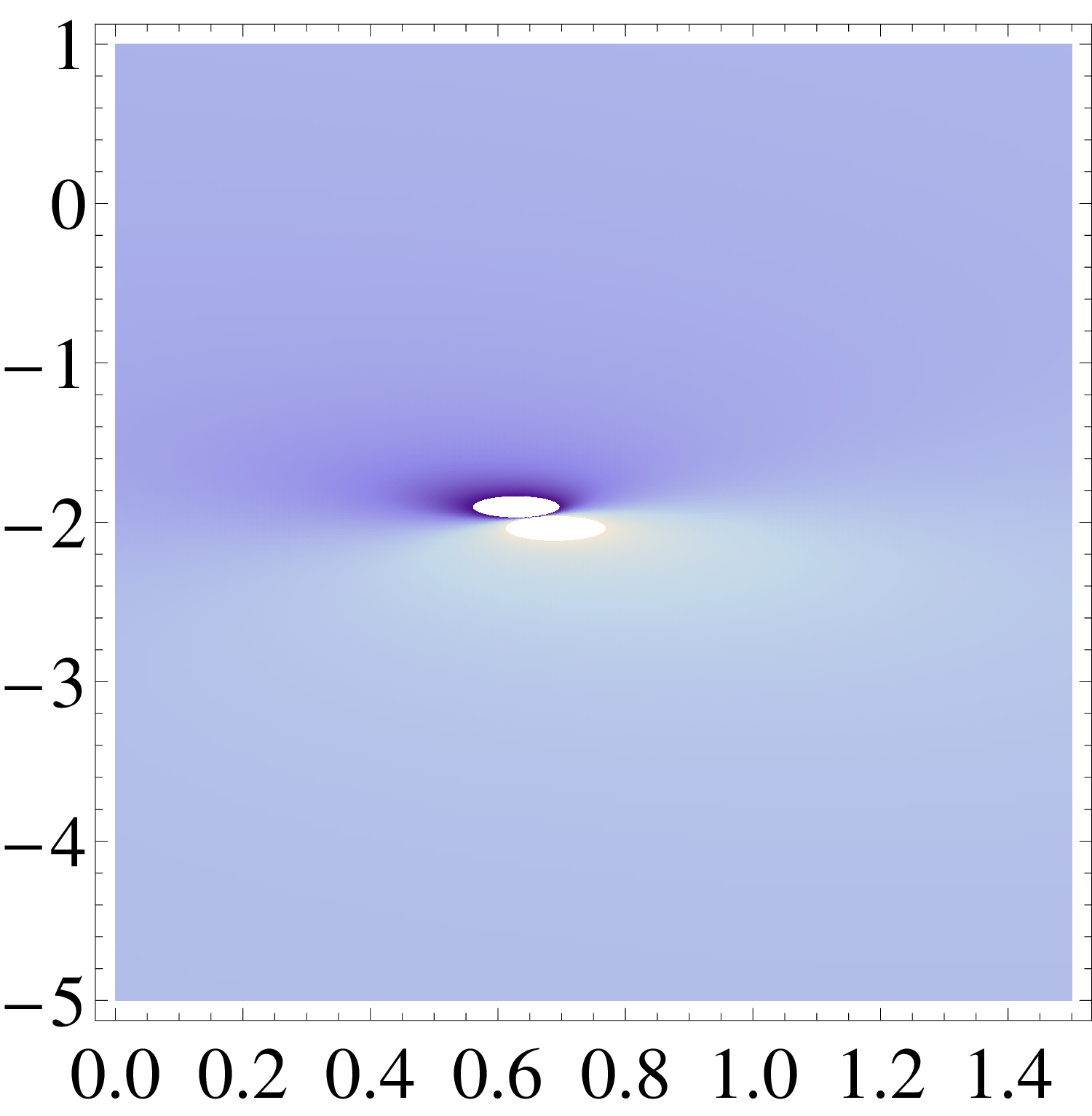}
\includegraphics[width=8.5cm]{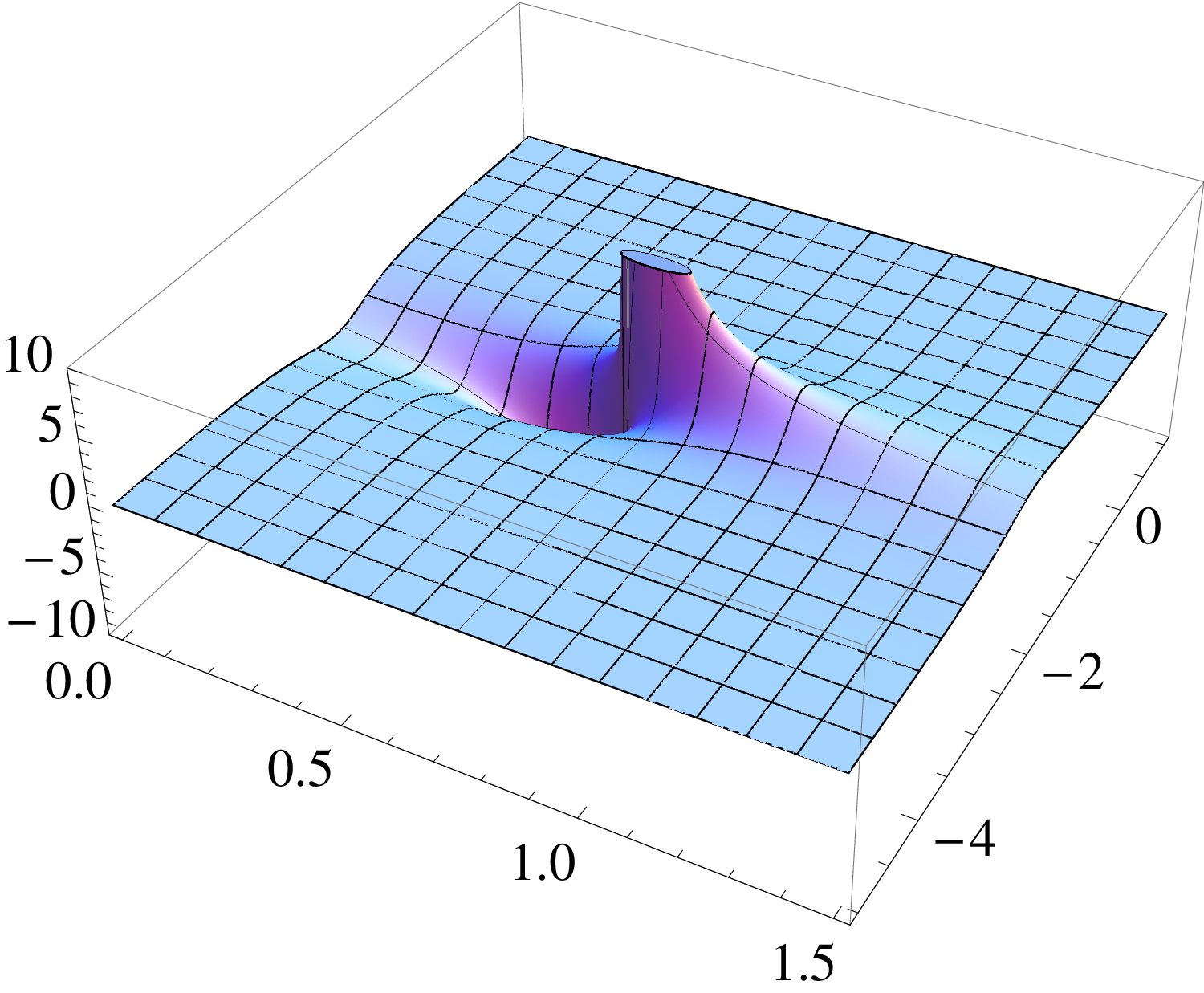}
\includegraphics[width=8.5cm]{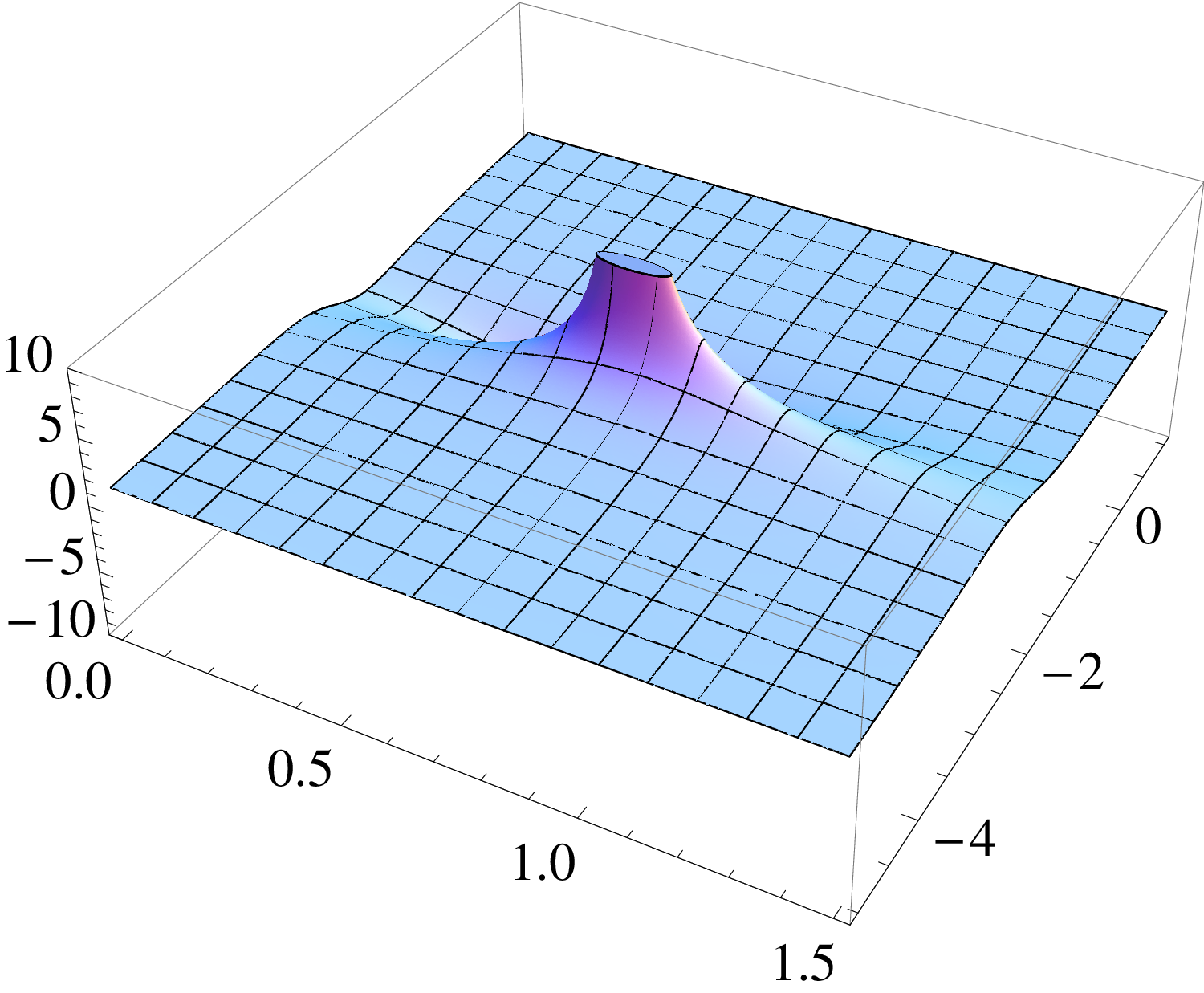}
\caption{\label{fig:polef4000} Same as in figure~\ref{fig:polef400}
but increasing the scale to $f=4$ TeV.
}
\end{figure}

Among the various reasons why a low-energy interaction can be strong is a pole in the scalar channel, as the $f_0(500)$ or $\sigma$ now well established in $\pi\pi$ scattering. Such pole  of the $t_{w}$ amplitude in the second Riemann sheet is mandated by a zero of Eq.~(\ref{zeroposition}). We have found such scalar pole, that is clearly visible in Figures
~\ref{fig:polef400} and~\ref{fig:polef4000}.


\begin{figure}
\centerline{\includegraphics[width=8.5cm]{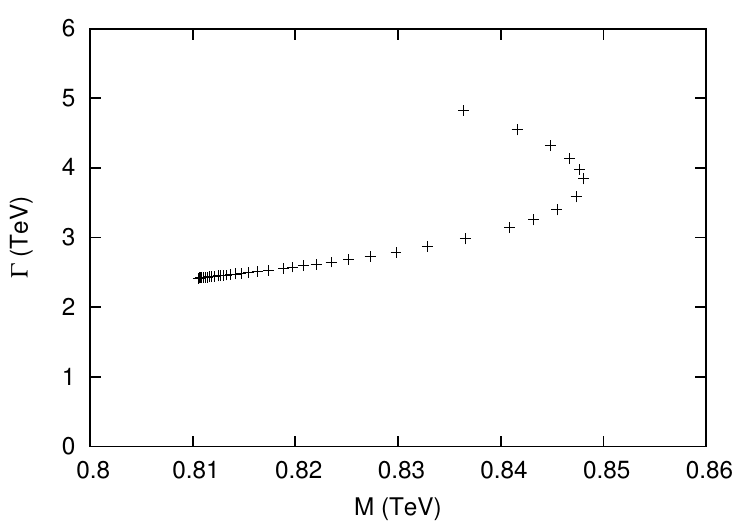}}
\caption{\label{fig:polemotion}
Motion of the pole position $s_0=(M^2,-M\Gamma)$ of $|t_\omega|$ from Eq.~(\ref{unitscattering}) in the complex $s$ plane for $f\in(250{\rm GeV},6{\rm TeV})$. For large $f$ the mass of the very broad ``resonance'' stabilizes
at about $810$ GeV, within reach of the LHC, but the width remains huge and the net effect is probably just an enhanced $W_LW_L$ amplitude in the $s$-wave.}
\end{figure}

To complete the analysis, we have followed the motion of the pole at $s_0$ in the complex $s$ plane. If the pole admits a resonance interpretation (a dubious case for such broad structures as we find), then one can interpret
$\Real(s)=M^2$, $\Imag (s) = - M \Gamma$. We plot such ``mass'' and ``width'' in Figure~\ref{fig:polemotion} for various values of $f$.

The pole position is seen to stabilize, for $f\gg v$, around 810 GeV. The width is so broad (above 2 TeV) that one can hardly talk of a resonance. Thus, although within reach of the LHC for large values $f\gg v=246$ GeV, the extraction of the pole from experimental data will have to be dispersive, by first extracting the amplitude $|t_\omega|$ and then prolonging $s$ to the complex plane; an attempt to obtain a Breit-Wigner resonant width from experimental data will yield very model-dependent results.

\begin{figure}
\centerline{\includegraphics[width=7cm]{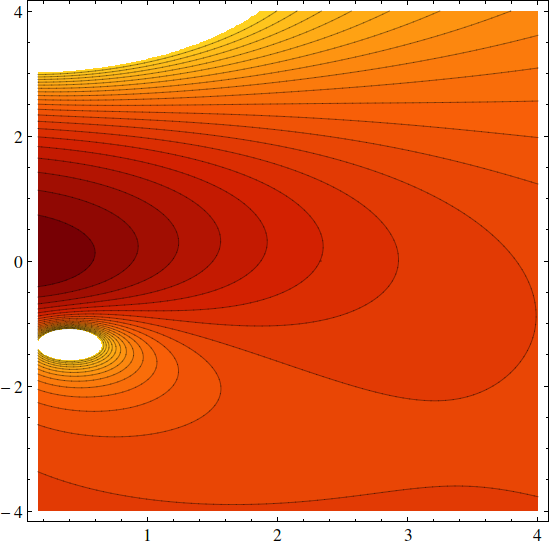}}
\caption{\label{fig:coupledchannelpole}
Modulus of the unitarized $t_\omega$ in a contour plot in the second Riemann sheet of  complex-$s$.
Plotted values range from 0 to 2 (never exceeding 1 for physical $s$). Note the characteristic pole below the real axis (white circle). Here $f=260$ GeV is very close to $v$, so the strong interaction comes entirely from the coupled channels ($\beta\ne \alpha^2$).
}
\end{figure}

A very interesting numeric exercise is to keep $\lambda_3$, $\lambda_4$ small as above, so that the $\varphi\varphi$ interaction is perturbative, make $f=260$ GeV $\simeq v$ so that $t_\omega$ is also perturbative for all the relevant low-energy range, but choose $\beta=3$ so that the interchannel coupling is strong.
Then we generate a strong, resonant interaction between the $\omega\omega$ and $\varphi\varphi$ channels that dynamically generates a so called ``coupled channel resonance'', a pole on the second Riemann sheet of Mandelstam's $s$ that is absent if we take $\beta \to \alpha^2=1$, but is there for larger coupling. This effect is clearly  visible in Figure~\ref{fig:coupledchannelpole}.  

\subsection{Large $N$}

\begin{figure}
\centerline{\includegraphics[width=8.5cm]{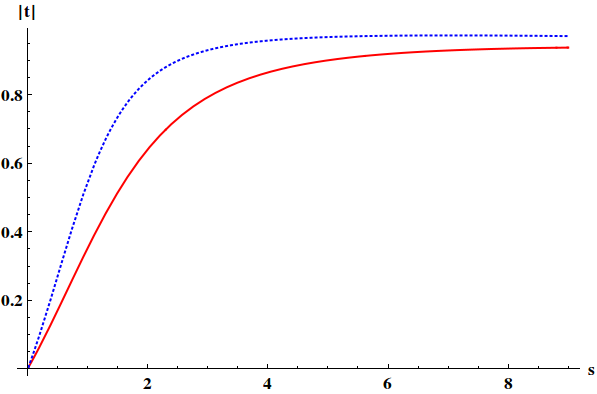}}
\caption{\label{fig:largeNcomp}
We compare the large-$N$ approximation (top, dashed line) with the K-matrix approximation (bottom, solid line), showing how the moduli of the amplitudes obtained are rather similar. In both cases only the elastic $\omega\omega\to \omega\omega$ was included, setting $\beta=1$ and $f=0.8$ TeV.}
\end{figure}

\begin{figure}
\centerline{\includegraphics[width=7cm]{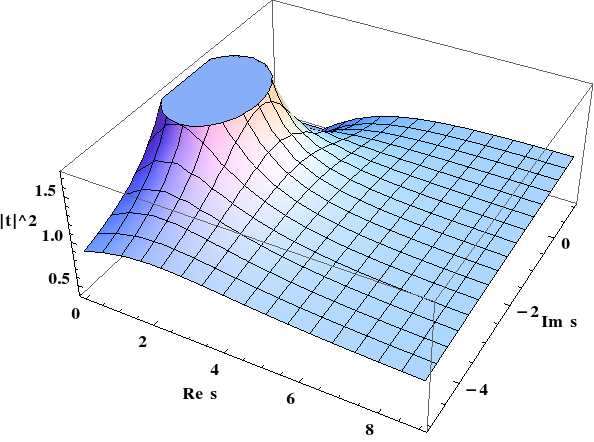}}
\caption{\label{fig:largeNpole}
Square modulus of the large-$N$ scattering amplitude $t_\omega$ in the second Riemann sheet of complex-$s$.
 Note the resonant pole below the real axis. The result is quite analogous to that of Figure
~\ref{fig:polef400} from the unitarized $K$ matrix amplitude.
}
\end{figure}

In Figure~\ref{fig:largeNcomp} we compare the result of the large-$N$ amplitude in Eq.~(\ref{fullLargeN}) with the K-matrix approach in Eq.~(\ref{unitscattering}) and see how comparable the two amplitudes are, in spite of the different mathematical expressions and derivation. Both approximations capture essentially the same physics of the $s$-channel cut.
Moreover in Figure~\ref{fig:largeNpole} we show the analytical prolongation of the large-$N$  amplitude to the second Riemann sheet where again we can see the above mentioned pole.

Numeric detail cannot be expected to be extremely precise, but neither does the actual experimental situation require it; for exploratory purposes it is very satisfactory that both approximations yield quite similar results in a strong-interaction scenario where perturbation theory is not applicable.

\subsection{Numerical treatment of the $N/D$ method}
\label{subsec:numericND}

\begin{figure}
\centerline{\includegraphics[width=7cm]{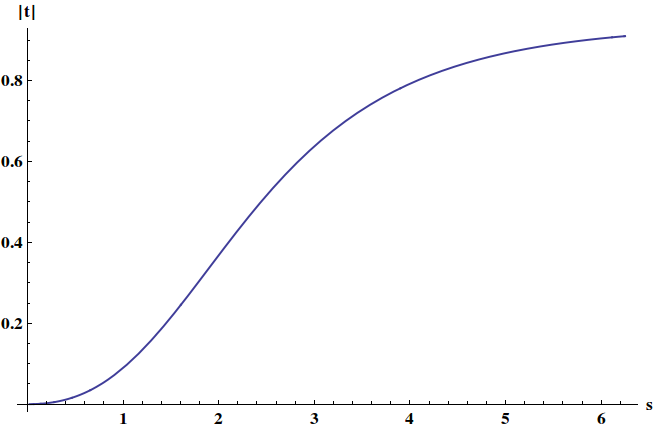}}
\caption{\label{fig:NDrealaxis}
Modulus of  $\tilde t_\omega$ in the $N/D$ method employing the analytical result of subsection~\ref{subsec:NoverD}. In this case we employ $f=1$ TeV,  $\beta=1$ and we have set the IR cutoff to $m=$ 150 GeV. 
}
\end{figure}

\begin{figure}
\centerline{\includegraphics[width=7cm]{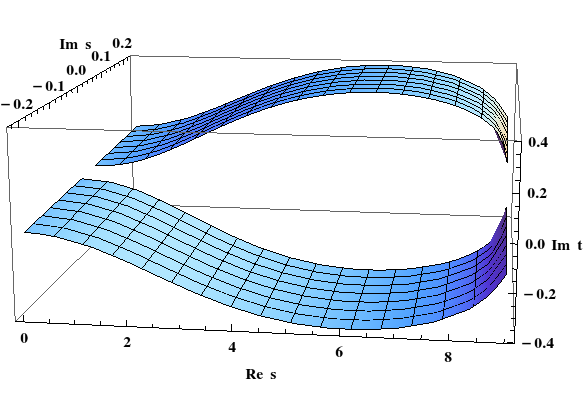}}
\caption{\label{fig:NDimaginary}
Imaginary part of the elastic amplitude  $\tilde t_\omega$ in the $N/D$ method employing the analytical result of subsection~\ref{subsec:NoverD}. Again, $f=1$ TeV, $\beta=1$ and $m=150$ GeV. This view shows the cut on the real $s$-axis
}
\end{figure}

In Figure~\ref{fig:NDrealaxis} we show the modulus of the $N/D$ unitarized $\tilde t^\omega$ in the one-channel case, put together from the explicit analytical solution of subsection~\ref{subsec:NoverD}.
The strength of the $\omega\omega$ scattering is apparent. The result is very similar to that found, for example, in Figure~\ref{fig:largeNcomp}, and again it is reassuring that the various sensible unitarization methods yield compatible results. It is also a nice check to look at the imaginary part of the full $N/D$ amplitude in Figure~\ref{fig:NDimaginary},
showing its characteristic discontinuity at the cut in the first Riemann sheet of the complex $s$ plane.

\begin{figure}
\centerline{\includegraphics*[width=8.5cm]{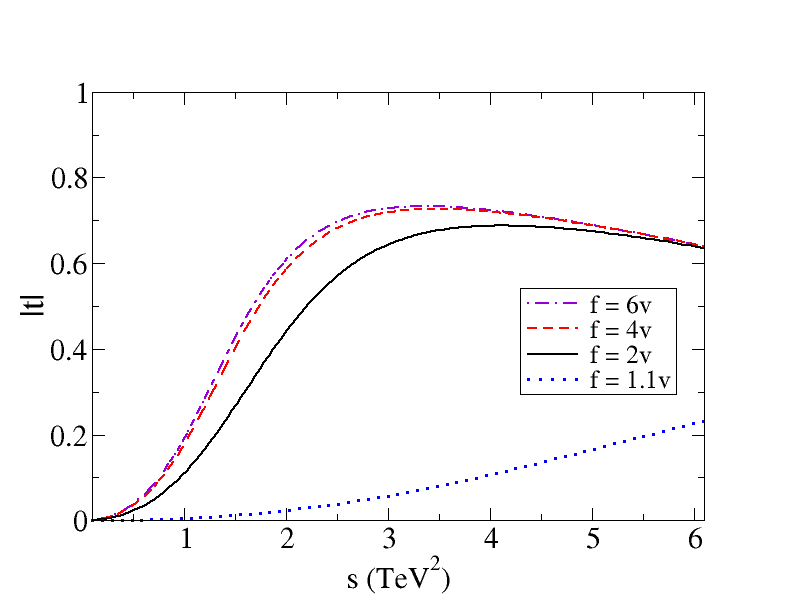}}
\caption{Dependence of $t_{00}$ in the $N/D$ method on the $f$ parameter (scaled by $v=246$ GeV). We have computed $N$ and $D$ numerically starting $Im (N)$  from Eq.~(\ref{unitarityinNoverD}) in perturbation theory.
 \label{fig:fdep}}  
\end{figure}

In Figure~\ref{fig:fdep} we show the modulus of the $\omega\omega $ amplitude as function of $s$ for three values of $f$ (in units of $v$) to show the dependence of the results with this unknown constant. 
The interactions become indeed strong although unitarity is not quite saturated (with our normalization this would correspond to $\ar t\ar=1$). Whether or not this happens is model-dependent.
It can be seen that, as $f\to v$, the amplitude is weaker and any structure recedes to higher energy. In the opposite limit, $f\gg v$, the maximum of the amplitude does not descend below the 1-2 TeV region. 
The result is in complete analogy to the behavior of the pole in the complex plane analyzed in Figure
~\ref{fig:polemotion}.

\subsection{The IAM method}

Obviously the numerical results coming from the IAM depend on the particular nature of the underlying dynamics through
the values of the renormalized EWCHL parameters $a_4(\mu)$ and $a_5(\mu)$ entering in $A(\mu)$. In particular the IAM unitarized partial wave can be written as:
\be\label{IAMpole}
\tilde t^\omega =\frac{ks}{1+\frac{k s}{\pi}\left[\log\frac{-s}{\mu^2}-\frac{\pi}{k^2}\left(A(\mu)+D\log\frac{s}{\mu^2}\right)\right]   }
\ee
as it is very easy to check provided
\be
k=\frac{1-\xi \alpha^2}{16 \pi v^2}=\frac{f^2-\alpha^2 v^2}{16\pi v^2 f^2}
\ee
or, in other words $t_0^\omega= k s$. We can relate the IAM to the K-matrix single-channel amplitude based on $t_0^\omega$, that was:
\be \label{MickeyMouse}
\tilde t^\omega =\frac{ks}{1+\frac{k s}{\pi}\left(\log\frac{-s}{\mu^2}+\log\frac{\mu^2}{\Lambda^2}\right)}
\ee
where we have introduced the arbitrary scale $\mu$.
This result is obviously simpler than Eq.~(\ref{IAMpole}) and in particular it does not show any LC. However it could be formally considered a particular case of the IAM result with $D=0$ and:
\be
A(\mu)=\frac{k^2}{\pi}\log\frac{\Lambda^2}{\mu^2}.
\ee
Therefore the IAM method improves the simple K-matrix by including the LC and also a contribution to $A(\mu)$ which depends on the dynamics underlying the EWSB. Thus the possible pole appearing in the second Riemann sheet 
by using the K-matrix will be modified correspondingly. 

In order to roughly understand the behavior of this pole we can set its position as $s_0=M^2-i M \Gamma$ and 
by choosing $\mu=M$ we get:
\be
M^2=\frac{k}{A(M)}=\frac{16 \pi^2 v^2 f^2}{(f^2-v^2)\log\frac{\Lambda^2}{M^2}}
\ee

\begin{figure}
\centerline{\includegraphics*[width=8.5cm]{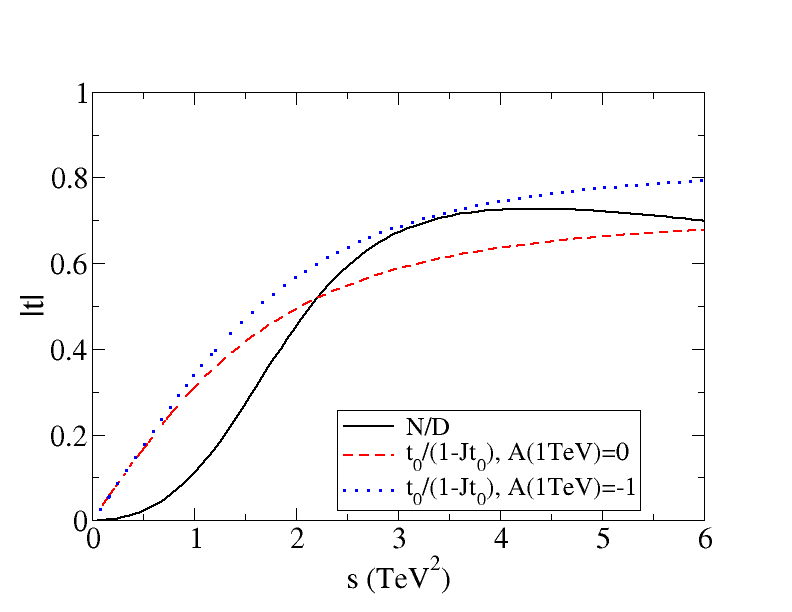}}
\caption{
Comparison between the $N/D$ method and a parametrization with a pole as in Eq.~(\ref{MickeyMouse}) that captures much of the same physics.
\label{fig:comp}
}
\end{figure}

Thus with the IAM we capture the features seen in the plot of the $N/D$ amplitude. As $f\to v$ its denominator vanishes and the pole position moves to $s=\infty$. But for large $f$, the pole position becomes saturated and constant, somewhat above 800 GeV, not accessing arbitrarily low energies, and thus avoiding nominal LHC exclusion bounds (though, for such a broad structure, a peak in the spectrum is not really expected).  In the opposite limit, $f\to v$, the pole position moves to large energy and essentially decouples from the low-energy spectrum. The interactions remain weak much longer.
A comparison between this toy parametrization and the $N/D$ calculation can be seen in Figure~\ref{fig:comp} for two values of $A$ (a negative quantity).

\section{Discussion and summary \label{sec:conclusions}}

In this article we take as phenomenological input the experimental situation after the first LHC run, namely, the existence of a Higgs-like light scalar resonance
(around 125 GeV) and the absence of any other states until about 600 GeV.  Therefore, the only light states associated to the Electroweak Symmetry Breaking Sector of the Standard Model are the would-be Goldstone bosons responsible for the electroweak boson masses, and this new light scalar resonance.

Extending the philosophy of the Electroweak Chiral Lagrangians, we have considered the most general low-energy effective Lagrangian with the appropriate symmetries including this new scalar. From this Lagrangian we have extracted the scattering amplitudes of those scalar bosons, that employing the Equivalence Theorem for energies $E\gg M_W\sim M_Z$ can be related with the electroweak longitudinal gauge boson scattering.

Excepting the minimal Standard Model case, these amplitudes 
grow linearly with $s$ (at relatively low energies $E\ll 4\pi v$)
and later violate unitarity badly.
In consequence we are forced to introduce some unitarization method. 
Since there is some controversy about which method is more appropriate (if any), we have explored several well known possibilities, including the K-matrix approach, large-N approximation, $N/D$ and the Inverse Amplitude Method. Qualitatively all these methods lead to similar results: 
a strongly interacting $s$-wave of $W_L W_L$ scattering,
and a pole of the amplitude in the second Riemann sheet.

In the particular case of the Standard Model there is no term linear in $s$ and the model is weakly interacting, renormalizable, and unitary by itself to very good accuracy at the perturbative level because of the lightness of the Higgs mass.

However it is very important to stress that in general the $W_LW_L$   interactions become strong for most parameter space. In particular we have found:
\begin{itemize}
\item For $\alpha^2 \ne \beta$ and $f > v$ we have strong $W_LW_L$ interactions, and strong coupling with the $\varphi\varphi$ channel.
\item For $\alpha^2=\beta$, $f>v$  strong elastic interactions are expected for $W_LW_L$, and a second, broad scalar analogous to the $\sigma$ in nuclear physics possibly appears. We identify a pole at 800 GeV or above in the second Riemann sheet very clearly, the question is whether it corresponds to a physical particle since it is so broad~\footnote{
A second scalar particle $S$ beyond the ``Higgs'' $\varphi$ induces two additional terms in the renormalizable $\mathcal{L}$.
For a doublet $S$, the mass Lagrangian 
$ a \varphi^\da \varphi + b (S^\da \varphi + \varphi^\da S) + c S^\da S $
can have  a negative eigenvalue for small $a\simeq 0$,
 $E\simeq \frac{1}{2} (b-\sqrt{b^2+c^2})$, a possible mechanism for the otherwise ad-hoc Mexican-hat potential. For a singlet $S$, 
$ \lambda \ar \varphi\ar^2 S^2 $, quartic in the fields, can also lead to negative mass term for $\varphi$ if $S$ acquires a vacuum expectation value (e.g., dilatation symmetry breaking) and $\lambda<0$,  acceptable if the quartic couplings of $S$ and $\varphi$ themselves are large enough.}.

\item Even if $f\simeq v$, with small $\lambda_i$, but we allow $\beta>\alpha^2$, one can have strong dynamics resonating between the $W_LW_L$ and $\varphi\varphi$ channels, likewise possibly generating a new scalar pole of the scattering amplitude in the sub-TeV region.
\item Finally, as an exception, for $\beta=\alpha^2=1$, $f=v$  we recover the Minimal Standard Model with a light Higgs which is weakly interacting.
\end{itemize}

Since the perturbative interactions grow with $s$, we have employed several unitarization templates based on the perturbative amplitudes. Unitarization models introduce a systematic error by concentrating on the right cut of the amplitudes and only approximating the left cut. Still, this is controlled by the fact that the amplitude is wanted for positive $s$, that is, on the upper lip of the RC. Let us then assess what uncertainty we should expect.

The LC is far in the complex plane, and suppressed by the $s'-s$ denominator that appears in the dispersion relation that ultimately underlies all these models (most clearly seen in the $N/D$ method). 

Our perturbative treatment of the LC is reliable to a typical $s\propto -(4 \pi f)^2$, so the relative error incurred on that left cut is not of order 1 before then. We are interested in the unitarized amplitude for $s \propto (4 \pi f)^2$ on the physical RC (for smaller $s$ there is no need to unitarize anything, the perturbative counting will do as well). The variation of $\tilde t^\omega$ on the RC has as typical scale $f$. We expect that the integral over the RC will be dominated by an interval of width $f$ around the given $s$ where the amplitude is wanted. Thus, we expect the unitarization methods to have a typical relative error $f/(2\times 4\pi f)\sim 1/(8\pi)$, or at the 5-10\% level. In order to  assess this systematic error we have employed different unitarization methods and found qualitatively similar results.

We have insisted that the pole we find in the second Riemann sheet of the WBGB scattering amplitude can only with difficulty  be interpreted as a physical state. This is because the structure is very broad and its main role is to make interactions strong. Other authors have reported new possible scalar states, for example in lattice calculations~\cite{Maas:2012tj} involving Higgs-Higgs dynamics.

Finally, it remains to address a philosophical question. If one takes as ultimate guiding principle the renormalizability of the theory, one finds  the MSM, with a very specific choice of parameters ($f=v$, $\alpha=\beta=1$) and higher energy scattering of longitudinal vector bosons will be weak.  Instead, if one adopts the effective theory way of thinking, where all possible low-energy couplings are allowed in the Lagrangian and need to be measured, then one finds that most likely scattering will be strong. 

Ultimately the observation by Wilson that only relevant operators remain in the low-energy theory after integrating out high-momentum shells, implies that whoever adopts the first, {\it renormalizable} point of view, is already committing to any new physics being at a very high scale.

Instead, he who adopts the second possibility, for example that dilatation invariance is broken by $f$ at a higher energy scale than electroweak invariance is broken by $v$, is implicitly assuming that some new physics (that we find to be strong interactions and possibly a second scalar resonance) is not too far in energy.

There is no current empirical information allowing us to choose one or another alternative, so at the present time it is largely a matter of personal taste. We look forward to further LHC data guiding theory.

\vspace{3cm}
\emph{
FJLE thanks the hospitality of the NEXT institute and the high energy group at the University of Southampton, ADG thanks the CERN-TH, and both the hospitality of the ECT* at Trento during the final stages of this work.    
This work was supported by spanish grant FPA2011-27853-C02-01
and by the grant BES-2012-056054 that supports RDL.}

\appendix

\section{Dispersive integrals \label{app:Integrals}}
In order to compute the dispersive integrals necessary for the algebraic treatment of the $N/D$ method in subsection~\ref{subsec:NoverD}, we
consider the more general IR and UV regularized integrals:
\be I=\int_{m^2}^{\Lambda^2}\frac{ds' f(s')}{s'-s-i\epsilon} \ee
where $s,s'\in (m^2,\Lambda^2)$ and $f(s')$ is analytic around this
interval. By using  the well known distribution identity: 
\be
\frac{1}{s'-s-i\epsilon}= PV\frac{1}{s'-s}+i\pi \delta(s'-s)
\ee
one has
\be \label{basicPV}
I = PV[I] + i\pi f(s)
\ee
where PV stands for Cauchy's  principal value part, i.e.
\ba
PV[I]=  \lim_{\epsilon\rightarrow 0}\left(\int_{m^2}^{s-\epsilon}\frac{ds'f(s')}{s'-s}  +   \int_{s+\epsilon}^{\Lambda}\frac{ds'f(s')}{s'-s}   \right) = \nonumber \\
g(\Lambda^2,s)-g(m^2,s)-\lim_{\epsilon\rightarrow 0}(g(s+\epsilon,s)-g(s-\epsilon,s))\ \
\ea
where $g(s',s)$ is a primitive of $f(s')/(s'-s)$, that is, 
\be \label{primitive}
\frac{\partial g(s',s)}{\partial s'}=\frac{f(s')}{s'-s}\ . 
\ee
Now, since by hypothesis $f$ is analytic in a domain surrounding $s$, it is possible to expand $f(s')$ around $s$ as:
\be
f(s')=f(s)+f'(s)(s'-s)+\frac{1}{2}f''(s)(s'-s)^2+...
\ee
yielding the Laurent series
\be
\frac{f(s')}{s'-s}= \frac{f(s)}{s'-s}+f'(s)+\frac{1}{2}f''(s)(s'-s)+...
\ee
and then we can choose $g(s',s)$ as:
\be
g(s',s)=f(s)\log(s'-s)+f'(s)(s'-s)+\frac{1}{4}f''(s)(s'-s)^2+...
\ee
Therefore:
\ba
PV[I]   = 
g(\Lambda^2,s)-g(m^2,s)-\lim_{\epsilon\rightarrow 0}\left[f(s)\log\frac{\epsilon}{-\epsilon}+f'(s)2\epsilon+\Or(\epsilon^3)\right] = \nonumber \\
 g(\Lambda^2,s)-g(m^2,s)-i \pi f(s) 
\ea 
and canceling $i\pi f(s)$ upon substitution in Eq.~(\ref{basicPV})
we finally obtain the simple relation
\be \label{resultI}
I=\int_{m^2}^{\Lambda^2}\frac{ds'
f(s')}{s'-s-i\epsilon}=g(\Lambda^2,s)-g(m^2,s) 
\ee 
for any $g(s',s)$ satisfying Eq.~(\ref{primitive}). By using this result it is not difficult to compute the integrals needed in subsection~\ref{subsec:NoverD}. Some of them involve the dilogarithm function $Li_2(z)$ defined for $\ar z\ar<1 $ as
\be 
Li_2(z)=\sum_{n=1}^{\infty}\frac{z^n}{n^2}
\ee
and extended analytically for other values of $z$ by the integral form
of that series
\be
Li_2(z)= -\int_0^z dt \frac{\log(1-t)}{t} \ . 
\ee 
Note the particular values $Li_2(0)=0$ and $Li_2(1)=\pi^2/6$. For very large $z$ we also recall the very useful asymptotic behavior 
\be 
Li_2 (z) \simeq
\frac{\pi^2}{6}-\frac{1}{2}\log^2(-z) \ .
\ee 

A key property of the family of $I$ integrals in equation~(\ref{resultI}) is that all of them
can be analytically extended to the whole complex plane. 
In particular, $Li_2(z)$ is analytic except for a cut on the positive real axis starting from $z=1$, so that
$\Imag Li_2(\lvert z \rvert + i\epsilon)= \pi \log \lvert z \rvert \theta 
(\lvert z \rvert - 1)$. Thus, Eq.~(\ref{resultI}) allows the explicit computation of the $N/D$ amplitude presented in subsection~\ref{subsec:NoverD} since the tree-level amplitudes are simple enough that the $g$-primitives can be found by inspection or with a simple symbolic manipulation program.
\section*{References}


\begin{thebibliography}{99}
\bibitem{footnote} The amplitude in Eq.~(\ref{LSalgebraic}) can be thought of as the resummation of the relativistic Lippmann-Schwinger series with the scattering calculated at one loop. Normally this would lead to integral expressions, but the purely algebraic expression in which $T$ and $J$ are factorized (no potentials inside the loop integrals) is a feature of effective theories with polynomial expansions, called on-shell factorization (the off-shell, non-factorizable parts are proportional to tree-level diagrams and can be reabsorbed in the parameters)~\cite{Oller}.
\end{thebibliography}
\end{document}